\documentclass[]{aa}
\pdfoutput=1
\usepackage{txfonts}
\usepackage{graphicx}
\usepackage{amssymb}
\usepackage{natbib}

\begin{document}

\title{The Search for Faint Radio Supernova Remnants in the Outer Galaxy:
Five New Discoveries}

\author{Stephanie Gerbrandt \inst{1,2} Tyler J. Foster \inst{2,3} \and Roland 
Kothes \inst{2} \and J\"orn Geisb\"usch \inst{2} \and Albert Tung \inst{1,2}}

\titlerunning{Five New Discoveries}
\authorrunning{Gerbrandt et al.}

\offprints{S. Gerbrandt, T. Foster or R. Kothes\\ \email{stephanie.gerbrandt@alumni.ubc.ca}
\email{fostert@brandonu.ca} \email{Roland.Kothes@nrc-cnrc.gc.ca}}

\institute{Department of Physics \& Astronomy,
	University of British Columbia,
	6224 Agricultural Road,
	Vancouver, B.C. V6T 1Z1 Canada\\
\and
	National Research Council, 
	Herzberg Programs in Astronomy \& Astrophysics,
	Dominion Radio Astrophysical Observatory,
	P.O. Box 248, Penticton BC, 
	V2A 6J9, Canada\\
\and 
	Department of Physics \& Astronomy,
	Brandon University,
	270-18th Street, Brandon, MB
	R7A 6A9, Canada\\
}

\date{}
\abstract {High resolution and sensitivity large-scale radio surveys of the 
Milky Way are critical in the discovery of very low surface brightness 
supernova remnants (SNRs), which may constitute a significant portion of the 
Galactic SNRs still unaccounted for (ostensibly the {}``Missing SNR problem'').
}
{The overall purpose here is to present the results of a systematic, deep
data-mining of the Canadian Galactic Plane Survey (CGPS) for faint, extended
non-thermal and polarized emission structures that are likely the shells of
uncatalogued supernova remnants.} 
{We examine 5$\times$5 degree mosaics from the entire 1420~MHz 
continuum and polarization dataset of the CGPS after removing unresolved 
{}``point'' sources and subsequently smoothing them. Newly revealed extended
emission objects are 
compared to similarly-prepared CGPS 408~MHz continuum mosaics, as well as to
source-removed mosaics from various existing radio surveys at 4.8~GHz, 2.7~GHz, 
and 327~MHz, to identify candidates with non-thermal emission characteristics. 
We integrate flux densities at each frequency to characterise the radio spectra 
behaviour of these candidates. We further look for mid- and high-frequency 
(1420~MHz, 4.8~GHz) ordered polarized emission from the limb brightened 
{}``shell''-like continuum features that the candidates sport. Finally, we use 
IR and optical maps to provide additional backing evidence.} 
{Here we present evidence that five new objects, identified as filling all or
some of the criteria above, are strong candidates for new SNRs. These five are 
designated by their Galactic coordinate names G108.5+11.0, G128.5+2.6, G149.5+3.2, 
G150.8+3.8, and G160.1$-$1.1. The radio spectrum of each is presented, 
highlighting their steepness which is characteristic of synchrotron radiation. 
CGPS 1420~MHz polarization data and 4.8~GHz polarization data also provide 
evidence that these objects are newly discovered SNRs. These discoveries 
represent a significant increase in the number of SNRs known in the outer 
Galaxy second quadrant of longitude (90$\degr$~$<$~$\ell$~$<$~180$\degr$), and 
suggests that deep mining of other current and future Milky Way surveys will 
find even more objects and help to reconcile the difference between expected 
numbers of Galactic SNRs and the smaller number of currently known SNRs.} {}


\maketitle

\section{Introduction}
The presence of supernovae in the Milky Way Galaxy represents one of our most 
critical keys to understanding the evolution of our Galaxy's interstellar 
medium (ISM). Supernovae are the most significant source of chemical enrichment 
in the ISM and the resulting high velocities of their remnants ensure 
far-reaching distribution of heavy elements. As such, supernova remnants 
(SNRs) are both the source of and the means by which metal-enriched material is 
distributed throughout the Galaxy. Shock waves from supernova explosions in 
the Galaxy produce hot ionized tunnels in the ISM, sweep up an ever-increasing 
collection of dust and gas, ionizing the gas and bending and warping the 
Galactic magnetic field in the process. Supernova shock waves compress gas 
clouds triggering star formation, and generate a vast amount of cosmic rays 
that play an important role in pressure balance in the Milky Way ISM. 
\citet{padm01} estimates that more than half of the material in the Galaxy has
been processed by supernova remnants.


Therefore, to better our understanding of the ISM and our Galaxy, we 
require not only a detailed study of the individual physical properties of SNRs 
but also an accurate count of their numbers. If we assume a mean lifetime of 
radio shell-type SNRs of $\approx 60,000$~yrs \citep{frai94} and a supernova 
rate of one per 30-50 years in spiral galaxies like the Milky Way 
\citep{tamm94}, the Galactic ISM is expected to host one to two-thousand radio 
supernova remnants at any given time. Models of the angular distribution of 
Galactic \ion{H}{ii} regions and SNRs also predict about 1000 Galactic radio 
SNRs \citep{li91}. However, in 2009 only 274 confirmed radio SNRs were 
catalogued in the Milky Way \citep[][]{gree09}, with a few more added since 
\citep[e.g.][see catalogue of \citet{ferr12} for a somewhat more up-to-date 
list]{koth14,fost13,roy13,gao11}.
But there are still many {}``missing'' SNRs, likely due (in part) to 
difficulties in identifying low surface brightness objects that may be confused 
by the emission of other Galactic sources and compact background objects, or
due to the appearance of very distant SNRs as non-conspiuous unresolved 
{}``point''-like sources in current radio surveys of the Milky Way.

The Canadian Galactic Plane Survey (CGPS) 1420~MHz continuum data has an 
unprecedented spatial dynamic range \citep{tayl03}, allowing for detection of 
never-before-seen faint emission. Since 2001, the CGPS dataset has spawned the 
discovery of 10 new SNRs in the first, second, and third quadrants of the Milky 
Way: G85.4$+$0.7 and G85.9$-$0.6 \citep{koth01}; G107.5$-$1.5 \citep{koth03}; 
G96.0$+$2.0 and G113.0$+$0.2 \citep{koth05}; G108.2$-$0.6 \citep{tian07}; 
G151.2$+$2.9 \citep{kert07}; G152.4$-$2.1 and G190.9$-$2.2 \citep{fost13}, and 
very recently G141.2$+$5.0 \citep[a pulsar wind nebula,][]{koth14}. This paper 
presents evidence for the addition of five new faint SNRs to this growing list, 
and outlines their spectral and continuum properties.   


\section{Survey Data}

\subsection{21~cm \& 74~cm}
Both the $\lambda$21~cm Stokes I,Q,U continuum and $\lambda$74~cm Stokes I continuum data 
were obtained from the Canadian Galactic Plane Survey (CGPS) dataset, which is described 
in detail by \citet{tayl03}. The release of the CGPS 21~cm linear polarization data is 
described by \citet{land10}. The data were obtained with the Dominion Radio Astrophysical 
Observatory's Synthesis Telescope \citep[ST,][]{land00}. The radio continuum survey 
provides images of the ISM with a resolution of about 
$49\arcsec\times49\arcsec\textrm{csc}\left(\delta\right)$ for 1420~MHz and 
2$\farcm$8$\times$2$\farcm$8$\textrm{csc}\left(\delta\right)$ at 408~MHz. 
The three phases
of the CGPS cover a longitude range of 53$\degr$~$<$~$\ell$~$<$~193$\degr$ and a 
latitude range of $-3.6\degr < b < +5.6\degr$, with a high latitude extension 
in the  
range of 100$\degr$~$<$~$\ell$~$<$~117$\degr$, extending up to $b = +18\degr$. 
All data are made publicly available via the Canadian Astronomy 
Data Centre.


The calibration of the CGPS 74~cm data as described in \citet{tayl03} was 
improved using several extra catalogues,
as described in detail in a CGPS 408 MHz source catalogue paper
\citep{geis14} which is in preparation.

\subsection{11~cm}
The $\lambda$11~cm data were obtained from a radio continuum survey of the 
Galactic plane taken with the Effelsberg 100-m telescope. The survey at 
2695~MHz has a 4$\farcm$3 angular resolution and a sensitivity of 
$\Delta T_{B}=$50~mK \citep{furs90}.

\subsection{6~cm}
Data at $\nu = 4.8$~GHz (Stokes IQU) are from the Sino-German $\lambda$6~cm
polarization survey of the Galactic plane \citep{xiao11,gao10,sun07}, 
made with the Urumqi 25~m telescope, at a resolution of 9$\farcm$5. For the 
high latitude candidate G108.5$+$11.0, $\lambda$6cm data are taken from the 
4850~MHz survey data (GB6) made with the former Green Bank 91~m dish, with a 
resolution of 3$\farcm$7$\times$3$\farcm$3 \citep{con89}. Due to the method of 
observation, structures $>$20$\arcmin$ in declination are expected to be missed 
\citep{con89}.  

\subsection{92~cm}
Stokes I data for all candidates were taken from the Westerbork Northern Sky 
Survey \citep[WENSS,][]{reng97} at 327~MHz ($\lambda$92~cm) with 
54$\arcsec\times$54$\arcsec\textrm{csc}\left(\delta\right)$ resolution. The UV 
coverage of the survey provides sensitivity to structures from 1$\degr$ down to 
the resolution limit \citep{reng97}.

\subsection{Optical \& 60~$\mu$m Infrared}
Optical data was obtained from the Space Telescope Science Institute (STScI) 
Digitized Sky Survey (DSS2 Red) \citep{mcLe00}. The 60$\mu$m infrared data 
were obtained from the Improved Reprocessing of the IRAS 
(Infrared Astronomical Satellite) Survey (IRIS) \citep{mivi05}.

\begin{table*}
\begin{center}
\caption{Integrated flux and spectral properties of G108.5$+$11.0, 
G128.5$+$2.6, G149.5$+$3.2, G150.8$+$3.8, and G160.1$-$1.1 from point-source 
subtracted and smoothed maps at five radio frequencies. See Sec.~\ref{notes}
for discussion of the individual candidates and these properties.}
\label{flux}
\centering
\begin{tabular}{lr@{$\pm$}lr@{$\pm$}lr@{$\pm$}lr@{$\pm$}lr@{$\pm$}lr@{$\pm$}lc}
   \hline
   Flux (Stokes I)& 
   \multicolumn{2}{c}{G108.5$+$11.0} &
   \multicolumn{2}{c}{G128.5$+$2.6} &
   \multicolumn{2}{c}{G149.5$+$3.2} &
   \multicolumn{2}{c}{G150.8$+$3.8} &
   \multicolumn{2}{c}{G150-S-shell} &
   \multicolumn{2}{c}{G160.1$-$1.1} \\
\hline
   S$_{4812/4850}$[mJy] & 380 & 70 & 140 & 40 & 150 & 30 & 300 & 60 & 50 & 15 & 60 & 15\\
   S$_{2695}$[mJy] & \multicolumn{1}{c}{} & \multicolumn{1}{c}{} & 180 & 50 & 390 & 100 & 560 & 120 & 110 & 25 & 145 & 30\\
   S$_{1420}$[mJy] & 660 & 200 & 190 & 40 & 500 & 60 & 630 & 100 & 160 & 20 & 165 & 20\\
   S$_{408}$[mJy] & 1030 & 400 & 350 & 100 & 1060 & 110 & 800 & 150 & 310 & 40 & 615 & 100\\	
   S$_{327}$[mJy] & \multicolumn{1}{c}{} & \multicolumn{1}{c}{} & 530 & 180 & 1350 & 300 & 1200 & 380 & 370 & 85 & 845 & 185\\
\hline
   $\alpha$ (S$\propto\nu^{\alpha}$) & $-$0.41 & 0.22 & $-$0.44 & 0.08 & $-$0.71 & 0.08 & $-$0.38 & 0.10 & $-$0.62 & 0.07 & $-$0.94 & 0.10\\
   S$_{\nu=\textrm{1GHz}}$[mJy] & 734 & 24 & 255 & 20 & 590 & 44 & 665 & 64 & 185 & 11 & 265 & 24 \\
\hline
   Centre ($\ell,b$)& 
   \multicolumn{2}{c}{108$\fdg$51,$+$11$\fdg$04}&
   \multicolumn{2}{c}{128$\fdg$47,$+$2$\fdg$59}&
   \multicolumn{2}{c}{149$\fdg$48,$+$3$\fdg$20}&
   \multicolumn{2}{c}{150$\fdg$78,$+$3$\fdg$75}&
   \multicolumn{2}{c}{150$\fdg$55,$+$3$\fdg$46}&
   \multicolumn{2}{c}{160$\fdg$11,$-$1$\fdg$08}\\

   Size $\theta_{\textrm{maj}} \times \theta_{\textrm{min}}\times$angle& 
   \multicolumn{2}{c}{64$\farcm$9$\times$39$\farcm$0$\times$20$\degr$}&
   \multicolumn{2}{c}{39$\farcm$6$\times$21$\farcm$5$\times$30$\degr$}&
   \multicolumn{2}{c}{55$\farcm$6$\times$49$\farcm$3$\times$40$\degr$}&
   \multicolumn{2}{c}{64$\farcm$1$\times$18$\farcm$8$\times$50$\degr$}&
   \multicolumn{2}{c}{}&
   \multicolumn{2}{c}{35$\farcm$9$\times$13$\farcm$2$\times$41$\degr$}\\
\end{tabular}
\end{center}
\end{table*}

\section{Method}
To best represent extended continuum emission, unresolved {}``point'' 
sources in the 84 individual 21~cm CGPS mosaics (each 5$\times$5 degrees in 
size) are modelled and removed. We first locate these sources using a program 
in the DRAO export software package called \textit{findsrc}. To 
enhance the point-like sources, \textit{findsrc} examines an image using a 
matched point source wavelet filter. It then removes point-like source 
responses from the filtered image utilizing a method similar to a Clark-style 
''clean''. All point source components found are fed to a source box file which 
is passed to a program called \textit{fluxfit}, which then fits each point 
source with an elliptical Gaussian, and subsequently removes it. Occasional 
sources are missed by the automated passes, and each source-removed mosaic was 
subjected to a final visual search by eye to locate and manually model and 
subtract unremoved sources. Each final source-removed mosaic was then smoothed 
to a circular beam of FWHM 3$\farcm$8 in the UV plane using FFT techniques. 
Smoothing the source-removed map increases the signal-to-noise (S/N) ratio of 
extended emission while levelling the artefacts left by imperfect point-source 
subtraction. The final prepared mosaics contain S/N-enhanced shell-like or 
round, filled-centre extended structures which were previously undetected. 
After source removal and smoothing, mosaics were tiled together to produce a 
single source-removed 3.8-arcminute resolution supermosaic of the entire 
$\lambda$21~cm CGPS continuum survey. This supermosaic prepared as described 
above has 32768$\times$4096 19$\arcsec$ pixels, and is available as a single 
large FITS file from T. Foster and R. Kothes by request.

Candidate objects were sought by visual inspection of the prepared 
$\lambda$21~cm supermosaic. Many hidden structures such as crescents, complete 
or partial shells, filaments, and filled-centre objects that appeared round-ish 
or elliptical were identified in the smoothed, source-removed supermosaic. For 
each candidate we next make smaller individual 21~cm maps centred on them. 
Along with 21~cm, complementary maps from other radio surveys at 6~cm (Urumqi), 
11~cm (Effelsberg), 74~cm (CGPS) and 92~cm (WENSS) were also made, for a total 
of 5 maps at 5 frequencies. 21~cm contour boundaries are overlaid on the other 
frequency maps, and all point sources within them were removed. All maps at 
complementary frequencies were then smoothed to at least 3.8 arcminutes unless 
their original resolution was lower. By comparing the emission in all five 
frequencies we get an initial impression of the steepness of the radio 
spectrum, and can identify and separate thermal from non-thermal sources. For 
promising non-thermal candidates we then measure the integrated flux density 
and its uncertainty with the following method. For a given frequency, a polygon 
that encloses the emission of each object is defined by eye on the map, and the 
flux density within the polygon is measured. This same polygon defined at this 
reference frequency is then used to measure the integrated flux density in each 
of the other four frequencies. We then repeat this procedure with the other 
four frequencies as reference, until we have 5 individual integrated flux 
density measurements per frequency. For 21~cm and 74~cm maps, which have 
emission on spatial scales much larger than the candidates sizes, a twisted 
plane is fitted to the polygon's vertices to determine a model-background map 
which is subtracted. 

We then weight each flux density measurement by 1/error$^{2}$ and obtain the 
spectral index $\alpha_{5}$ for each candidate by fitting a power-law 
S$_{\nu}\propto\nu^{\alpha_5}$ to the five-frequency spectrum (note: only three 
frequencies were available for G108.5$+$11.0: $\lambda$6~cm, 21~cm, and 74~cm). 
The flux densities and spectral index for each SNR are tabulated in Table~1, 
including the flux density estimated at $\nu=$1~GHz.       

Errors in measurements reflect the variation in flux density $\sigma_S$ across 
the five different polygons. These errors were summed in quadrature with errors 
in the total intensity calibration $\sigma_c$, which are $\pm$4, 10, 5, 10 and 
20\% for $\lambda$6, 11, 21, 74, and 92~cm, respectively. For G108.5$+$11.0, 
calibration errors intrinsic to the 4850~MHz survey reference 2\% of the 
\citet{baar77} flux-density scale \citep{con89}, which indicates an absolute 
accuracy of about $\pm$5\% \citep{baar77}.

The angular size of each SNR (each $<$1$\degr$) is determined by measurements 
of the brightest emission regions (shells) by eye, and not extrapolated to the 
ill-determined full size of each faint candidate. Since the newly discovered
SNRs in this paper are all seen as incomplete shells that are confused with
unrelated Galactic foreground/background emission and \ion{H}{ii} regions, 
we do not estimate surface brightnesses. 

\section{Notes on Individual Objects}\label{notes}
We discuss each candidate in the context of our criteria for identification
of supernova remnants lurking within our extensive source-removed and smoothed 
21~cm mosaic. For each object, several criteria have been fulfilled, in order 
for the object to be considered a SNR. These criteria are (in order of 
importance): a non-thermal radio spectrum, a roundish limb-brightened (aka 
{}``shell''-like) appearance in several frequencies, presence of 6~cm and/or 
21~cm polarized emission within the continuum contours of the shell(s), the 
lack of correlated IR emission, and the possible appearance of filamentary 
optical emission.
 
In the following discussion of the individual candidates, the appearance and 
spectral features of each SNR candidate reference Figures~\ref{g108_maps} and 
\ref{g128_maps}-\ref{g160_maps}, presented mainly in the {}``cubehelix'' colour 
scheme of \citet{gree11}. All Stokes I radio maps are presented in 
their native resolution except as where noted in the text label above each 
figure panel, and point-sources in and around the vicinity of the candidate 
have been subtracted to emphasize the structure of the extended emission. 21~cm 
continuum contour levels overlaid on each of the maps in these figures are at 
the 4,5 and 6-sigma levels above the background (values given in the text label 
above each 21~cm I map). Spectra are presented in Figures~\ref{g108flux} and 
\ref{g128flux}-\ref{g160flux} with comparable $y$-axis scaling for direct 
comparison of their steepness'. We note that the polarization of an object's 
shell will appear different at the two wavelengths for which we have 
polarization data (6~cm and 21~cm). At $\lambda$6~cm Faraday rotation by the 
intervening ISM is low and polarized shell emission should be highly visible; 
however the resolution of these data is 9.5$\arcmin$, so that 
small-scale highly polarized emission such as is found in SNR shells is 
likely to be diluted away by foreground/background emission in the beam
solid angle. 
As well, 21~cm data presented here do not have short spacings added, and due to 
the 12$\times$ higher Faraday rotation as compared to 6~cm, 21~cm polarization 
structures in shells suffer much more from depolarization and confusion with 
foreground structures and also may not appear at all. In our discussion, 
North/South refer to the Galactic plane and not to the equatorial coordinate 
system. 21~cm Polarized Intensity (PI) maps shown in the corresponding figures 
are at 3.8-arcminute resolution, and PI maps at both frequencies are overlaid 
with \vec{E}-field vectors. 60$\mu$m IRIS (re-processed IRAS) maps 
presented are all scaled to the same brightness levels for relative 
comparison. 

\subsection{G108.5+11.0}
\begin{figure*}[!ht]
\centerline{
\includegraphics[bb = 118 242 506 575,height=5.4cm]{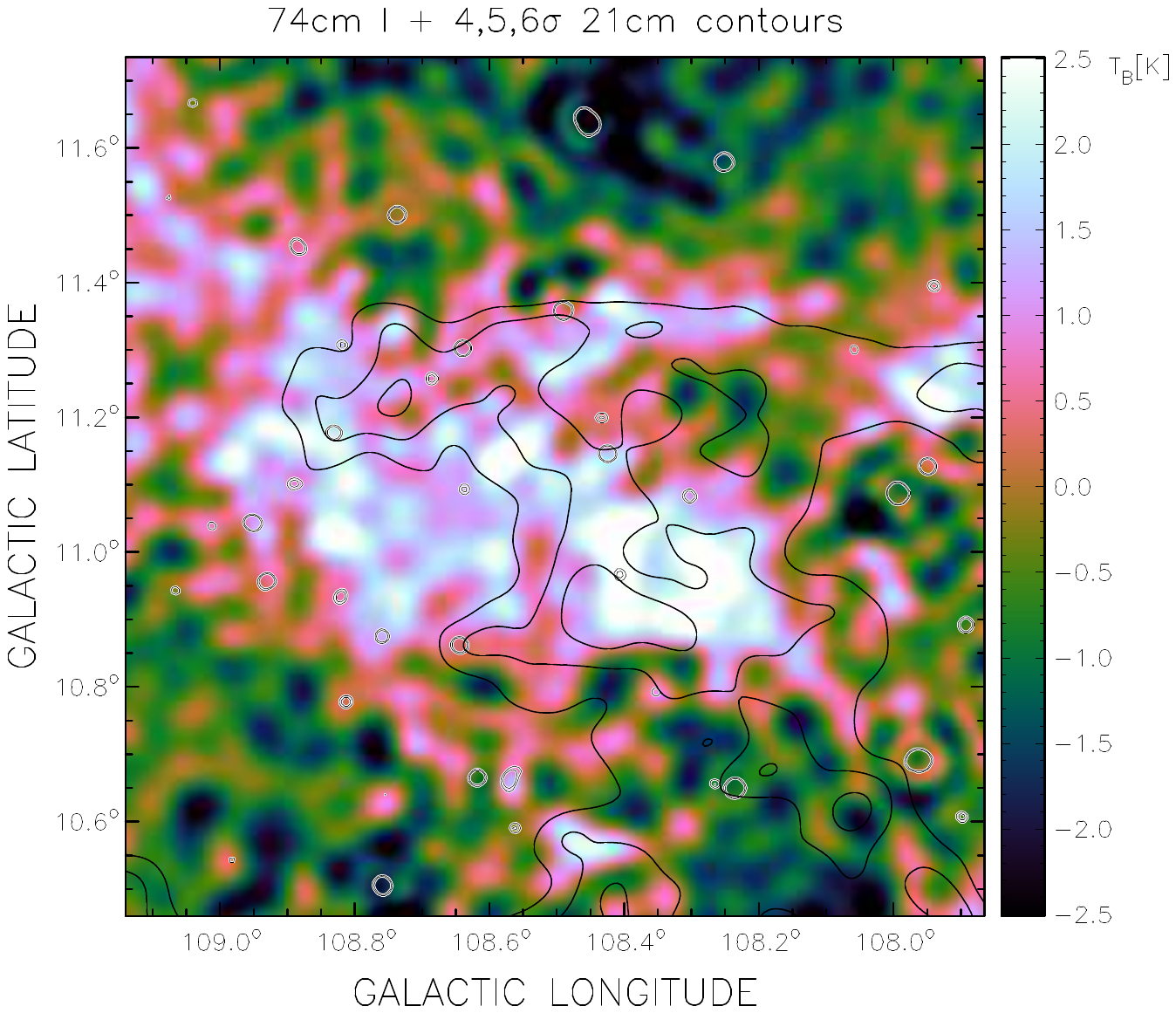}
\includegraphics[bb = 134 242 506 575,height=5.4cm]{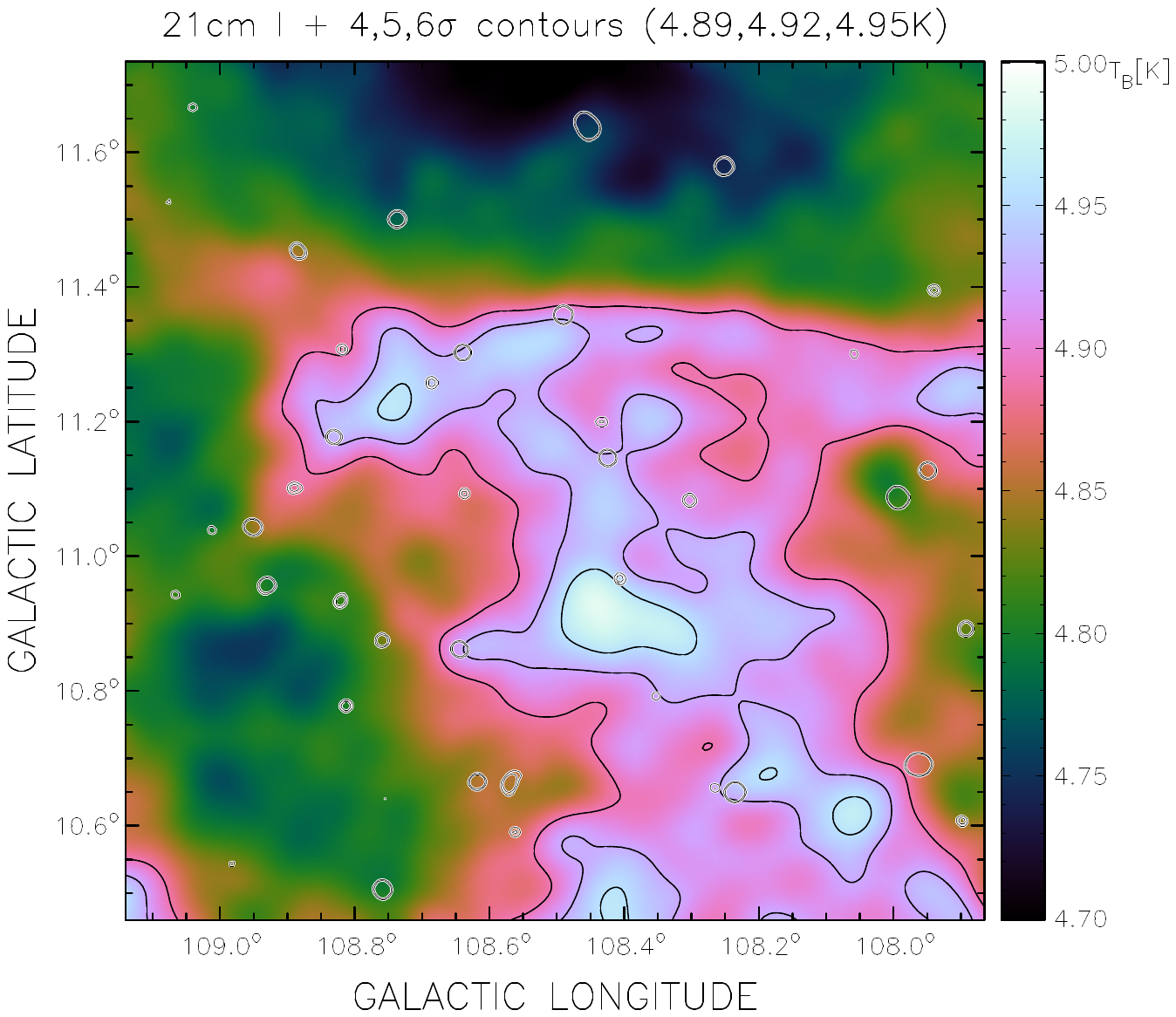}
\includegraphics[bb = 134 242 506 575,height=5.4cm]{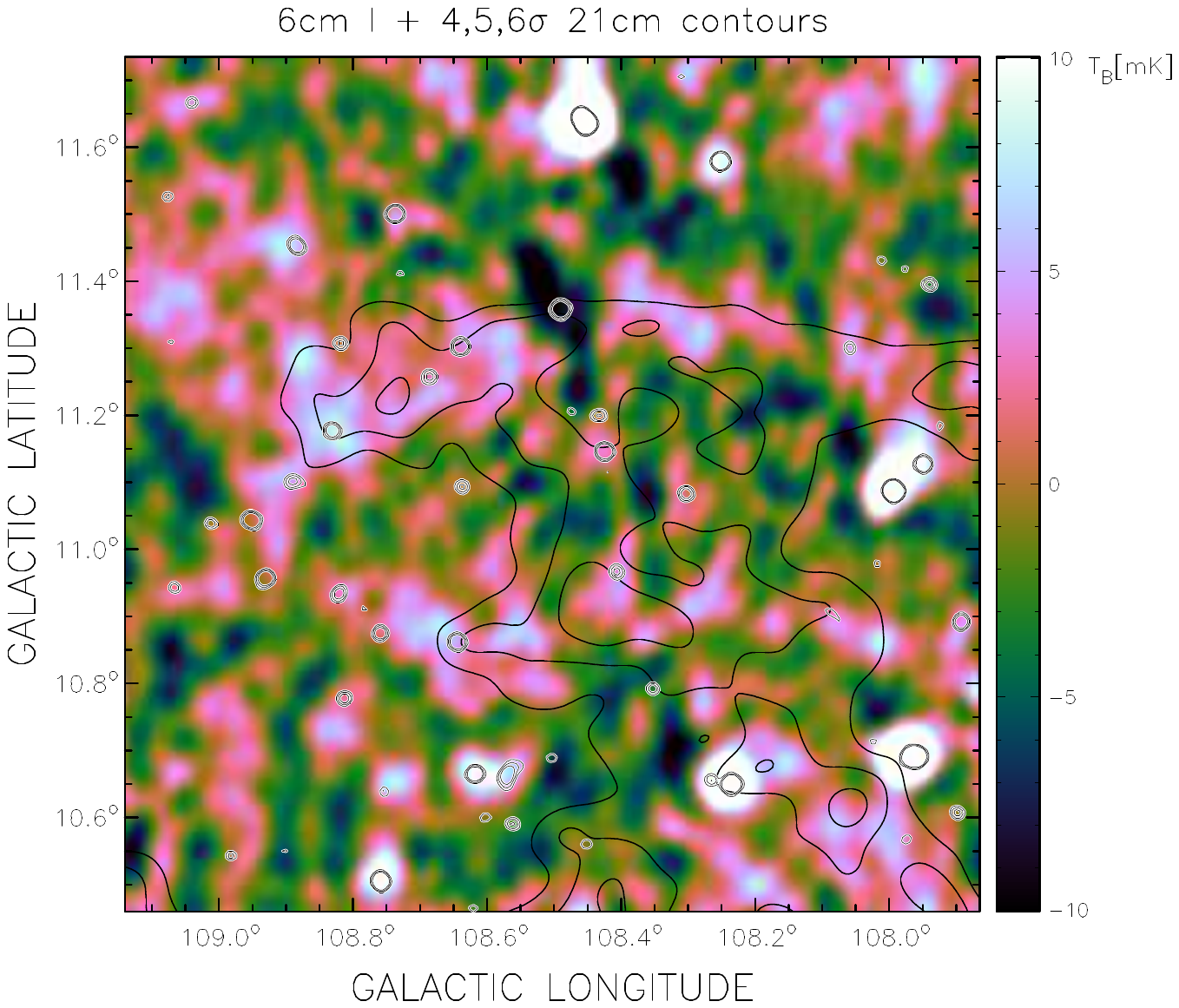}
}
\centerline{
\includegraphics[bb = 118 242 515 575,height=5.4cm]{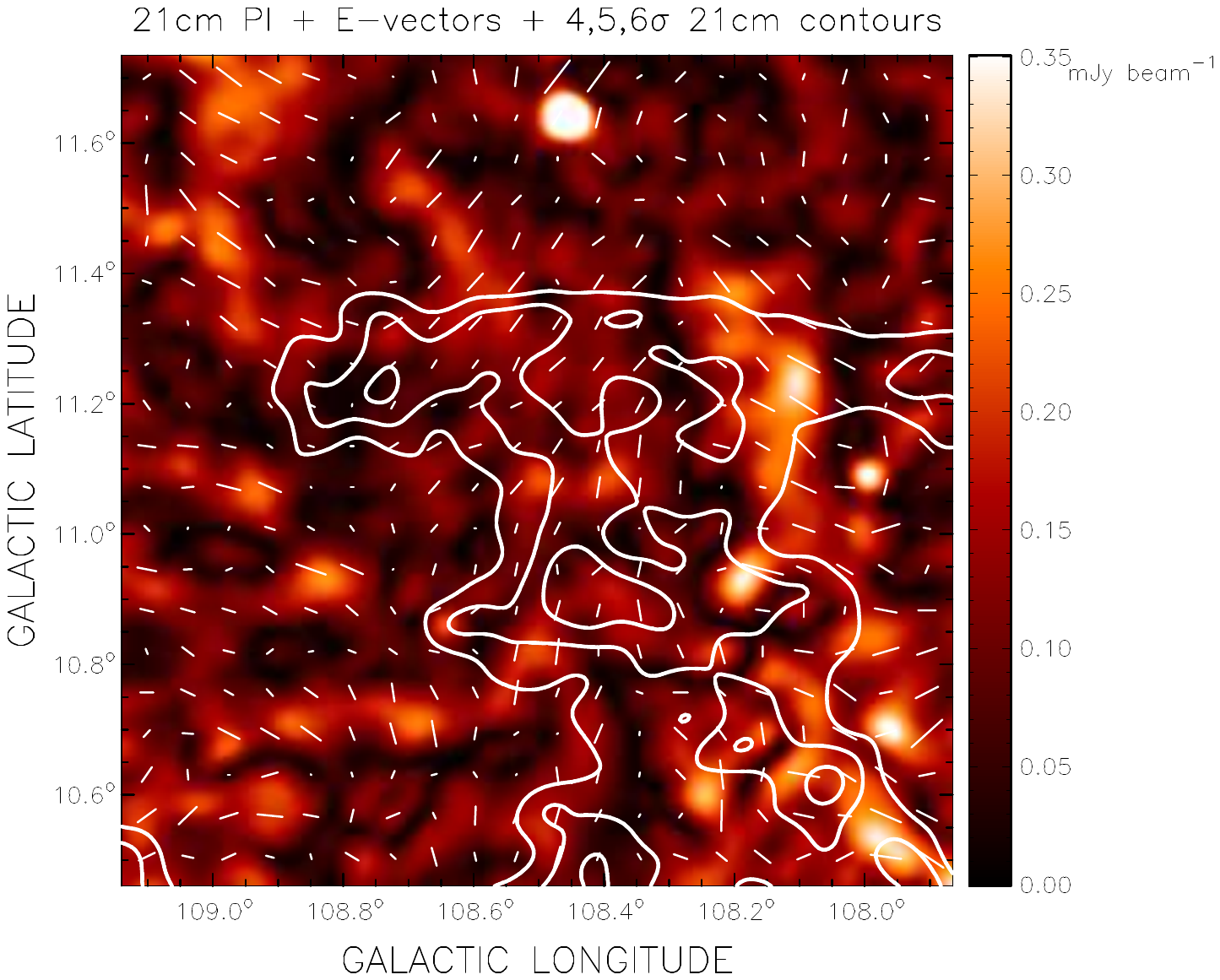}
\includegraphics[bb = 134 242 510 575,height=5.4cm]{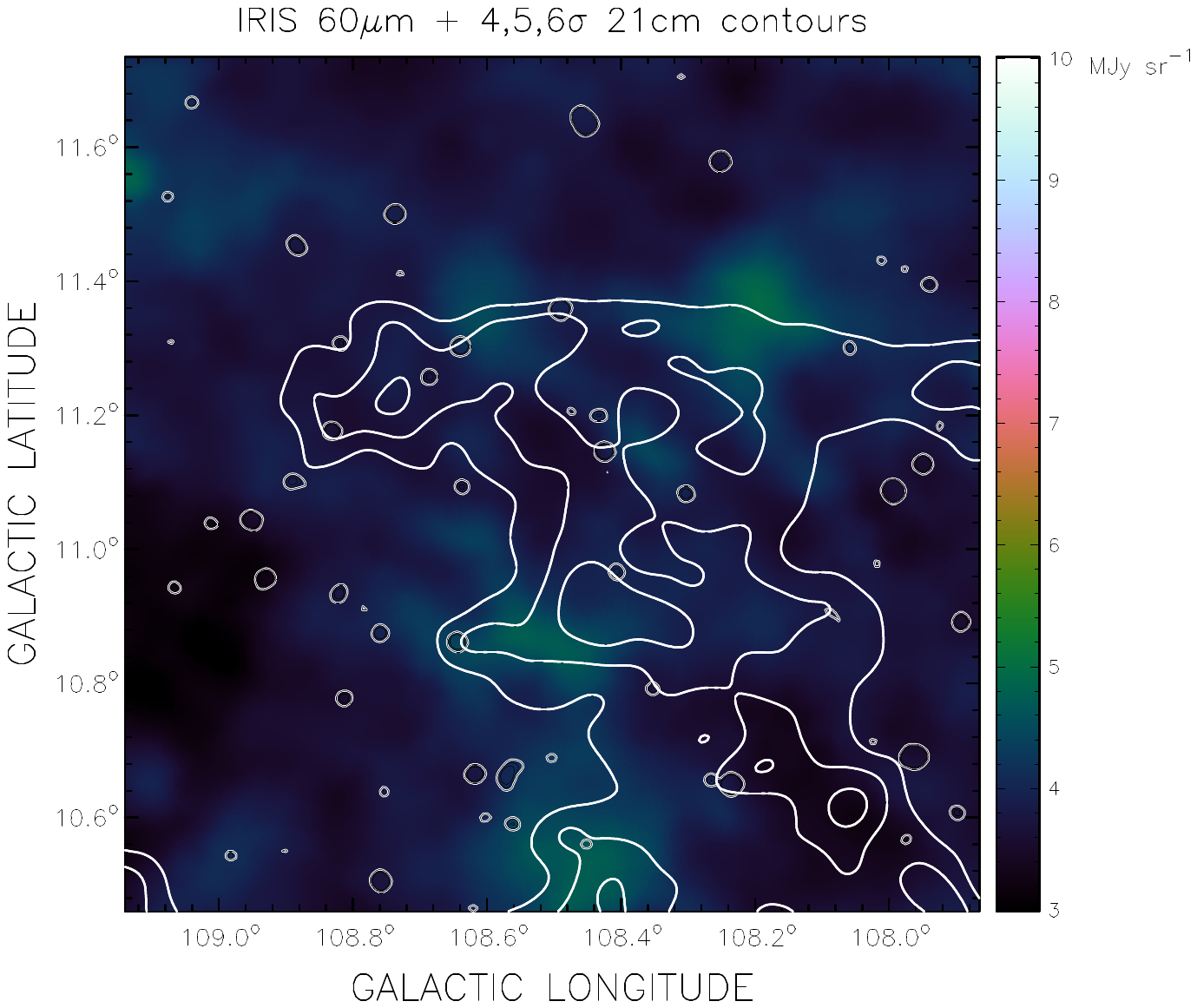}
\includegraphics[bb = 134 242 506 575,height=5.4cm]{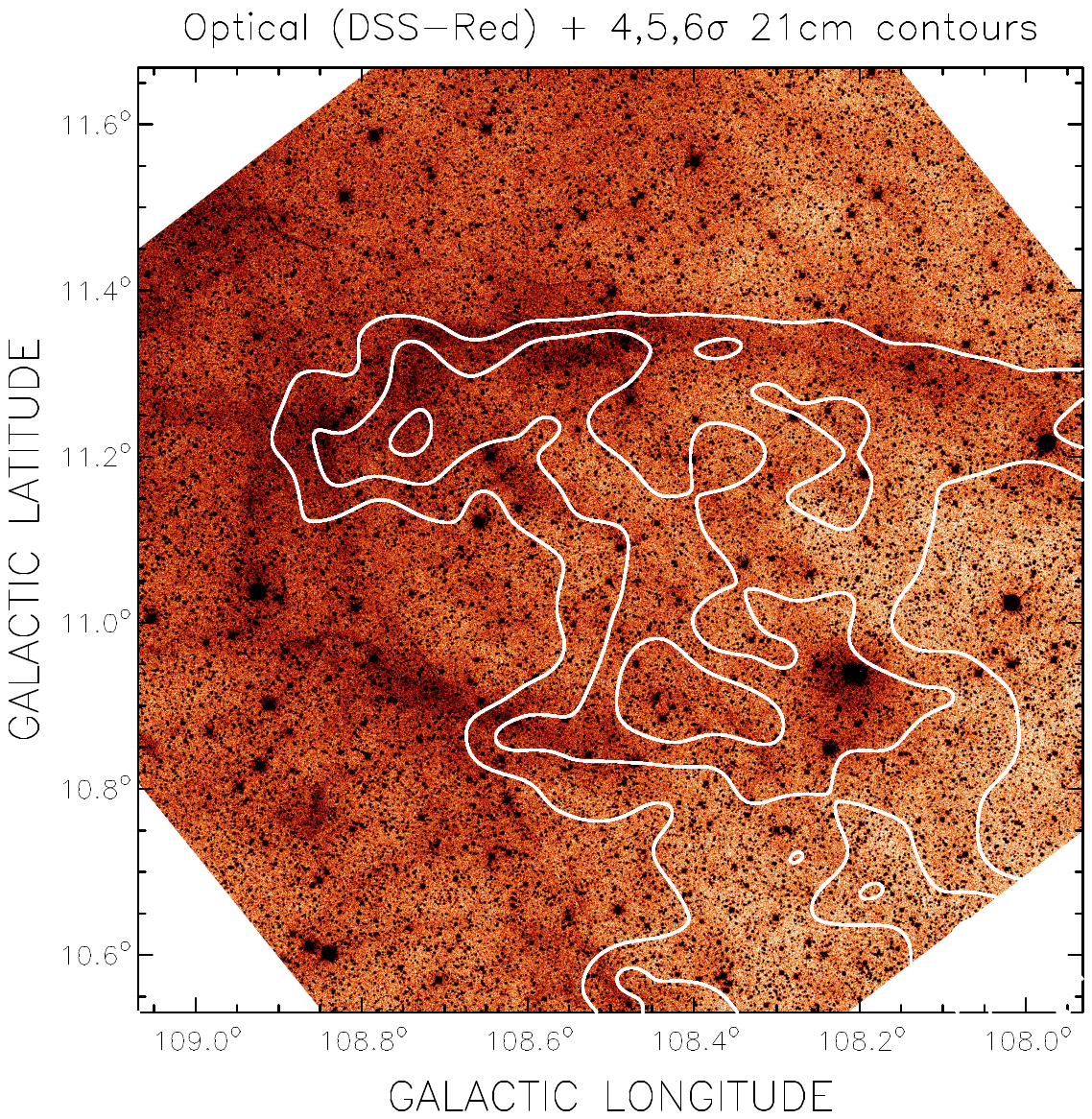}
}
\caption{\label{g108_maps} Total and polarized intensity maps of G108.5+11.0
at 408~MHz, 1420~MHz and 4.8~GHz (top right). The polarized intensity 
map at 1420~MHz is shown in the bottom left. An IRIS map of infrared emission 
at 60~$\mu$m (bottom centre) and a DSS2 (red) map (bottom right) covering the 
same region of the sky are shown for comparison. 
}
\end{figure*}

\begin{figure}[!ht]
\begin{center}
\includegraphics[scale=0.4,angle=0]{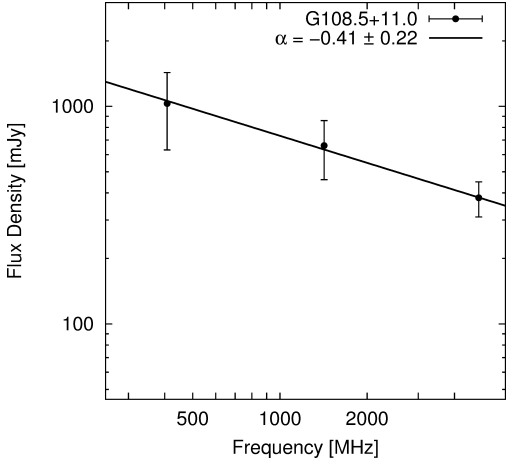}
\caption
{Integrated radio flux spectra for G108.5$+$11.0 using three frequencies and 
overlaid with a power law fit with error-weighted least squares.} 
\label{g108flux}
\end{center}
\end{figure}

Total power maps at 4.85~GHz, 1420~MHz, and 408~MHz are shown in 
Figure~\ref{g108_maps}. G108.5$+$11.0 is beyond the observed limits of the 
327~MHz survey data. The object has a prominent shell on top of a diffuse 
emission plateau. The integration area chosen for G108.5$+$11.0 included the 
well-defined Northern shell half and a less defined shell mixed with diffuse 
emission in the south. If these are both the top and bottom shell-halves of a 
single SNR, then G108.5$+$11.0 has an angular size of 
65$\arcmin\times$39$\arcmin$ and an angle between the bilateral symmetry axis 
of the SNR and the Galactic plane of $\sim$20$\degr$. G108.5$+$11.0 has a 
moderately steep spectral index of $\alpha_{3}=-$0.41$\pm$0.22, obtained with 
only three frequencies (see Figure~\ref{g108flux}; uncertainty from min-max 
slope). Due to missing large angular structures in the GB6 survey, we might 
expect a slightly lower integrated flux density measurement at $\lambda$6~cm, 
resulting in an artificially steeper spectrum using three frequencies. However, 
taking the $\lambda$21~cm and 74~cm data alone we obtain 
$\alpha_{74}^{21}=-$0.36$\pm$0.58 (min-max slope), similar to the 
three-frequency spectrum, albeit with a large uncertainty. 

The optical map (see Figure~\ref{g108_maps}) shows an elongated filament of red 
emission within the $\lambda$21~cm contours, with a well-delineated Northern 
edge tracing the North radio shell. The filament-like edge is indicative of an 
older SNR, which over time, has swept up increasing amounts of ISM dust and 
gas. As shown in Figure~\ref{g108_maps}, there are insignificant infrared 
emission patches in the field, none of which are spatially correlated with the 
well-defined N shell. 

No obvious 21~cm polarized emission is seen associated with the shells, but any 
such emission could easily be depolarized by Faraday rotation from an 
intervening foreground. G108.5$+$11.0 is not seen as a depolarizing object 
either, since there is very little 21~cm polarized emission in its field to 
compare to. Since G108.5$+$11.0 is a higher latitude remnant, there is no 
Urumqi $\lambda$6~cm polarization data available. Along with only three 
frequencies to determine its spectrum, the presence of optical emission and 
absence of infrared emission, we classify G108.5$+$11.0 as a good, but not 
strong candidate SNR. We hope that further, deeper observations will be 
performed to confirm our proposed classification of G108.5$+$11.0 as an SNR. 
Given its high latitude location and the very nice alignment of its N and S 
shells with the Galactic plane (and hence with the large-scale Galactic 
magnetic field), further study may also provide new insight to our 
understanding of SNRs and their evolution in the Galactic halo.

\subsection{G128.5+2.6}
\begin{figure*}[!ht]
\centerline{
\begin{minipage}{6.3cm}
\includegraphics[bb = 121 241 506 575,height=5.4cm]{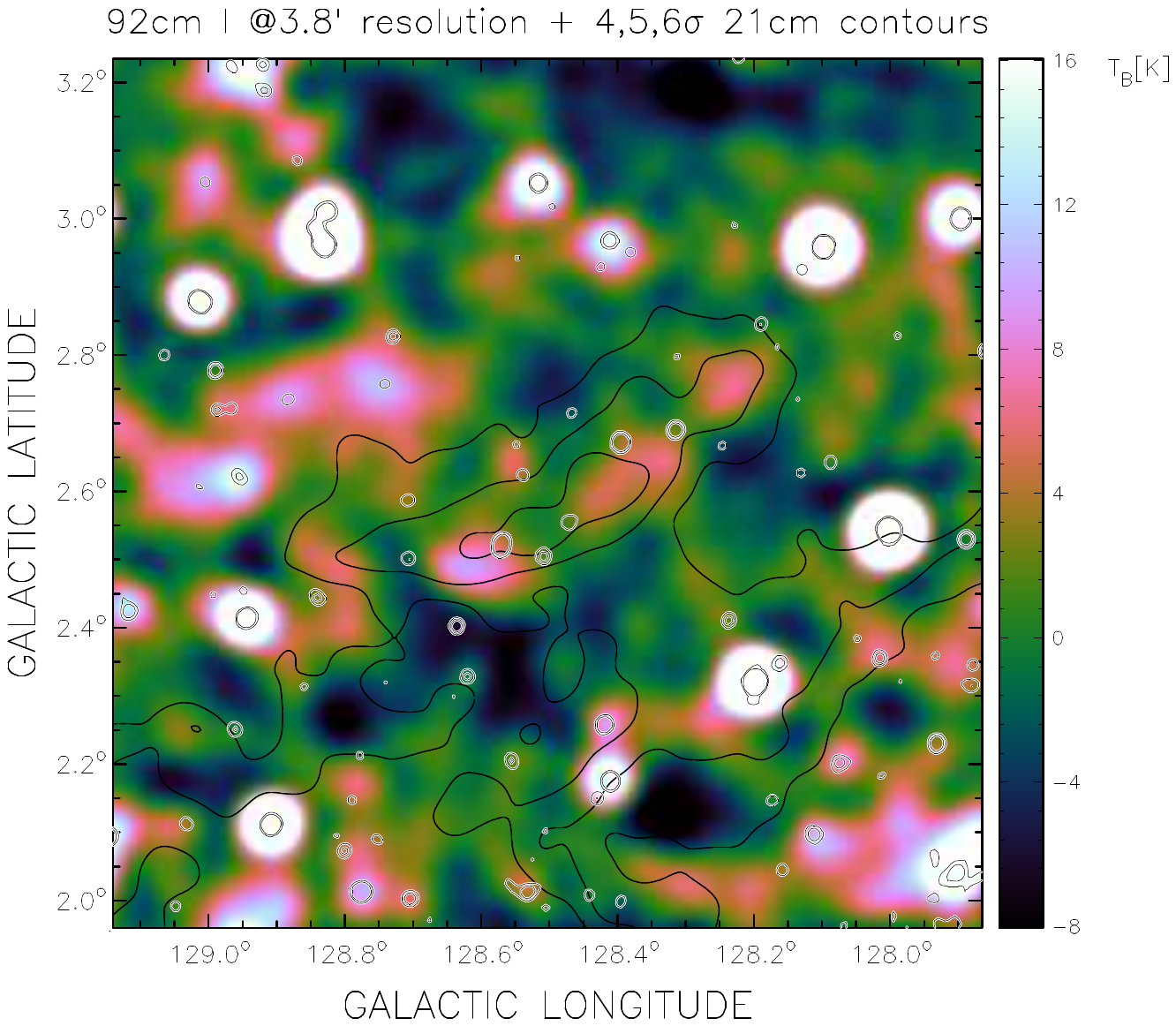}
\end{minipage}
\begin{minipage}{6.1cm}
\includegraphics[bb = 139 241 506 575,height=5.4cm]{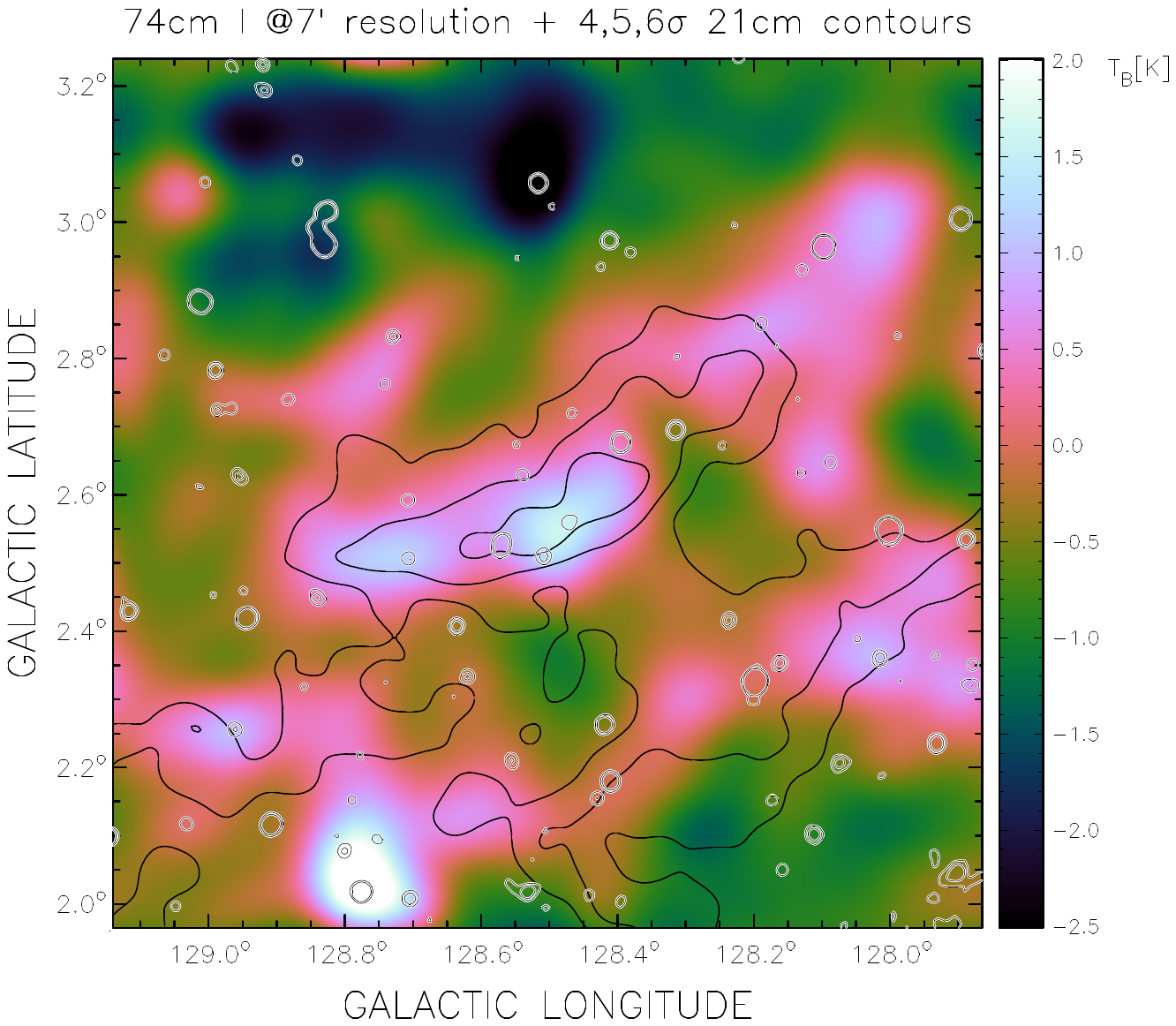}
\end{minipage}
\begin{minipage}{6.2cm}
\includegraphics[bb = 139 241 506 575,height=5.4cm]{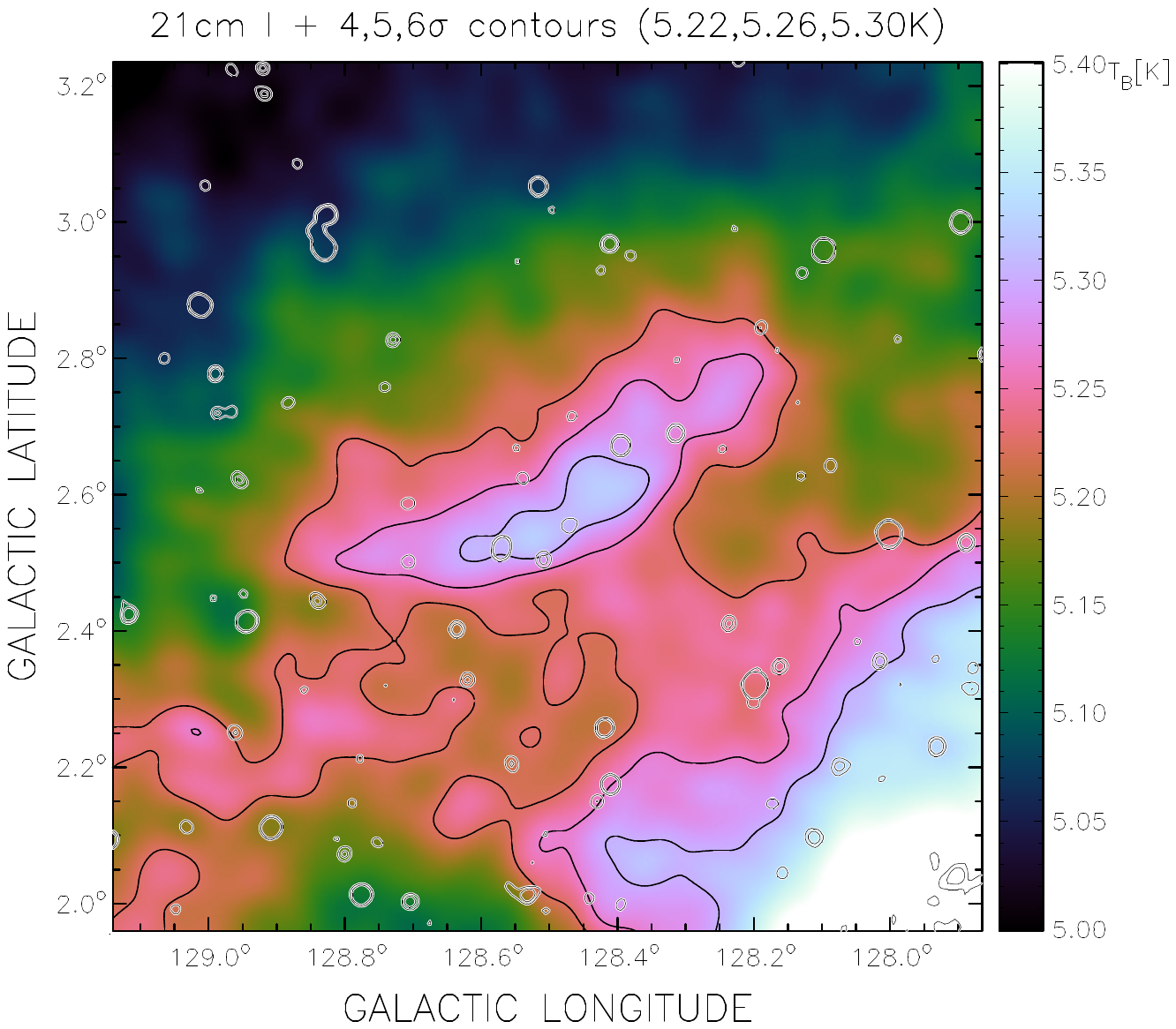}
\end{minipage}
}
\centerline{
\begin{minipage}{6.3cm}
\includegraphics[bb = 121 241 506 575,height=5.4cm]{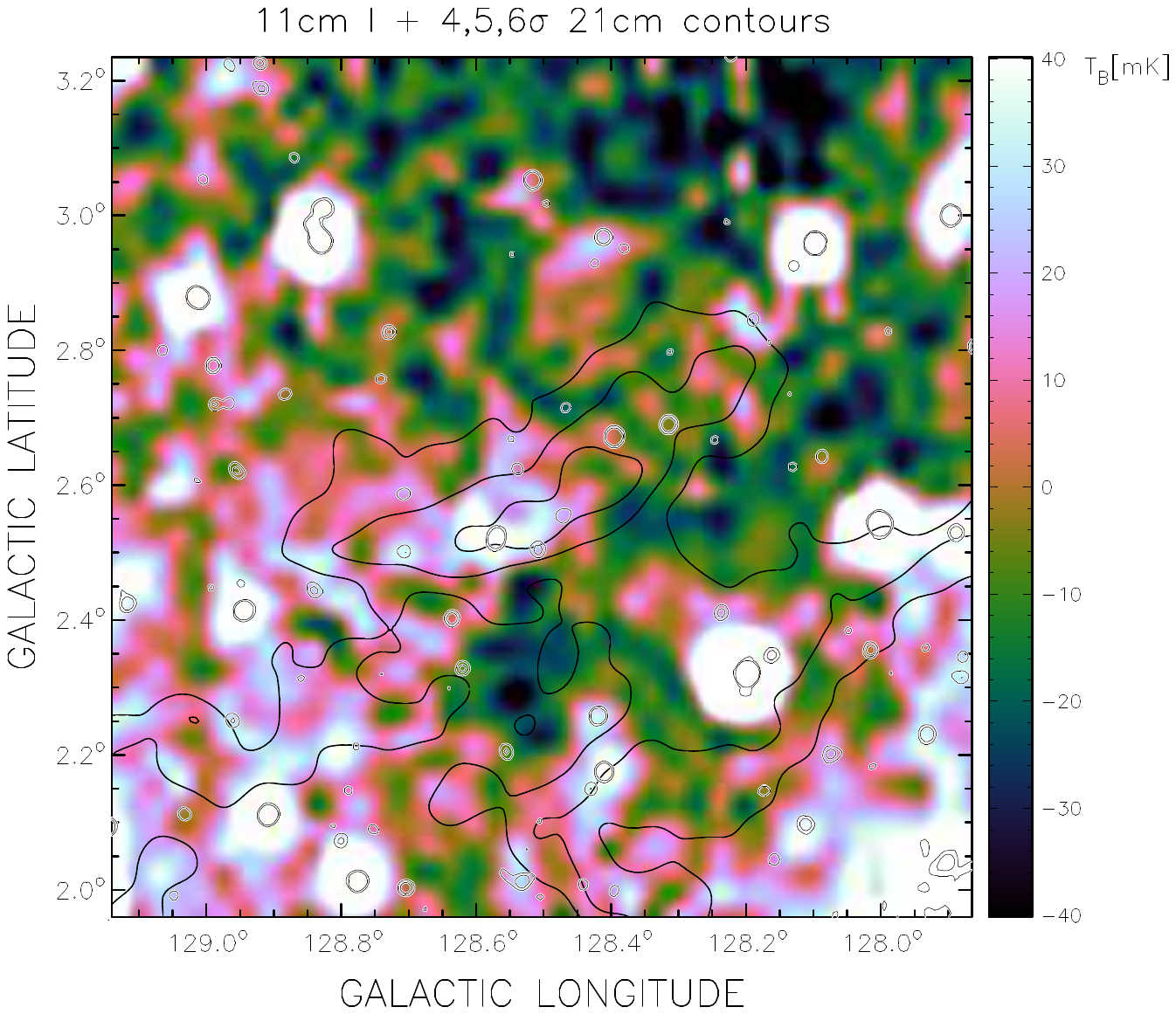}
\end{minipage}
\begin{minipage}{6.1cm}
\includegraphics[bb = 139 241 506 575,height=5.4cm]{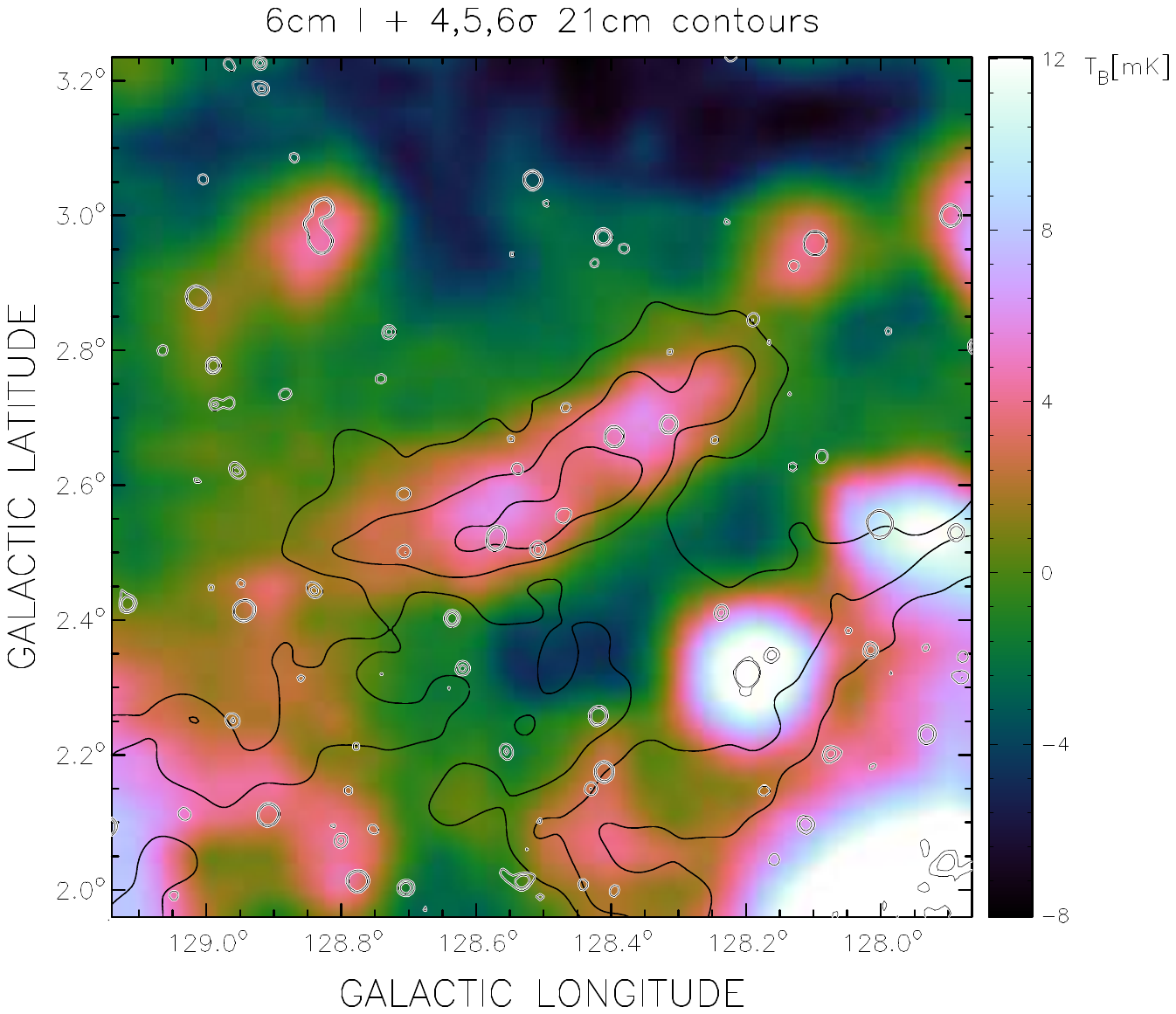}
\end{minipage}
\begin{minipage}{6.2cm}
\includegraphics[bb = 139 241 506 575,height=5.4cm]{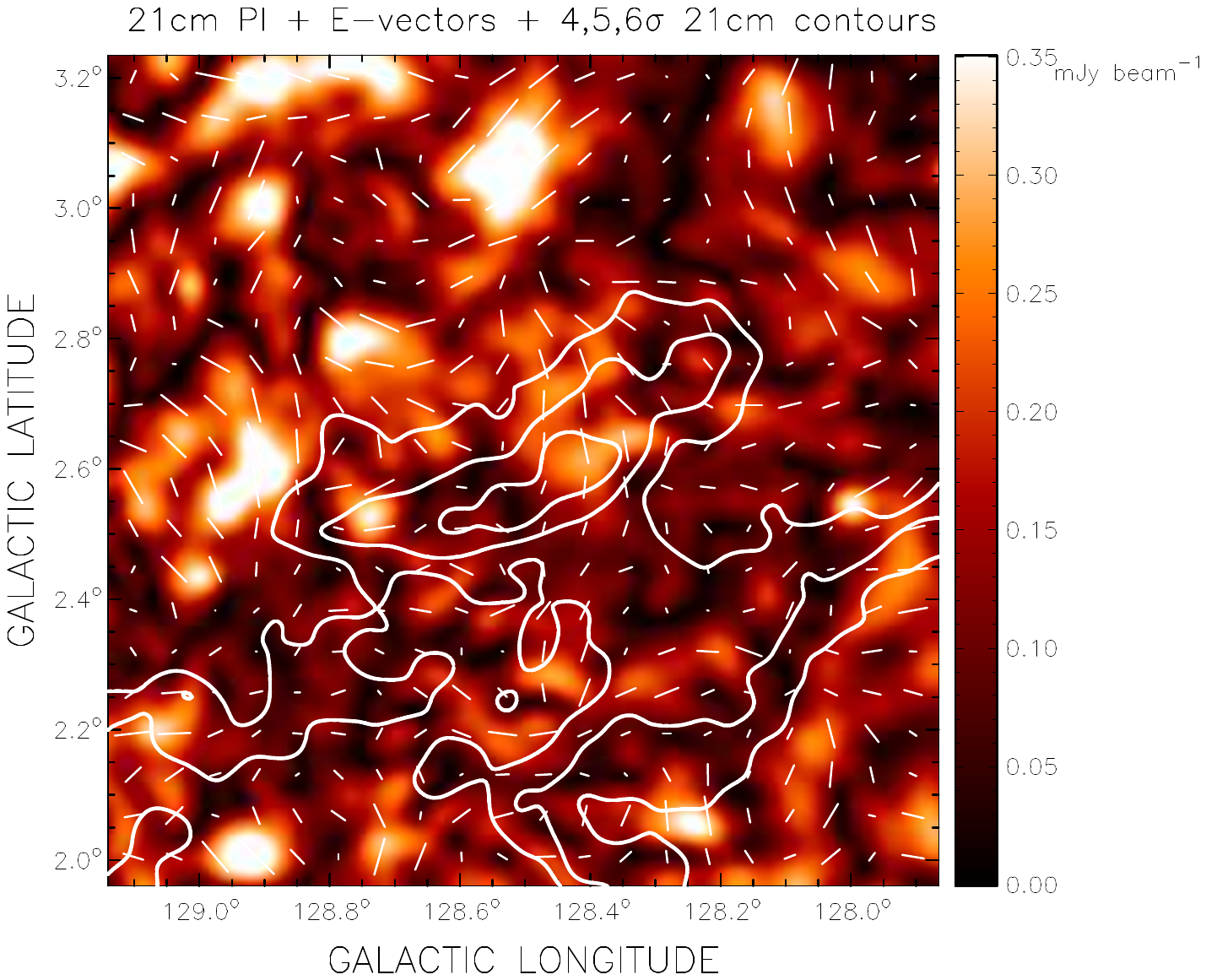}
\end{minipage}
}
\centerline{
\begin{minipage}{6.3cm}
\includegraphics[bb = 121 241 506 575,height=5.4cm]{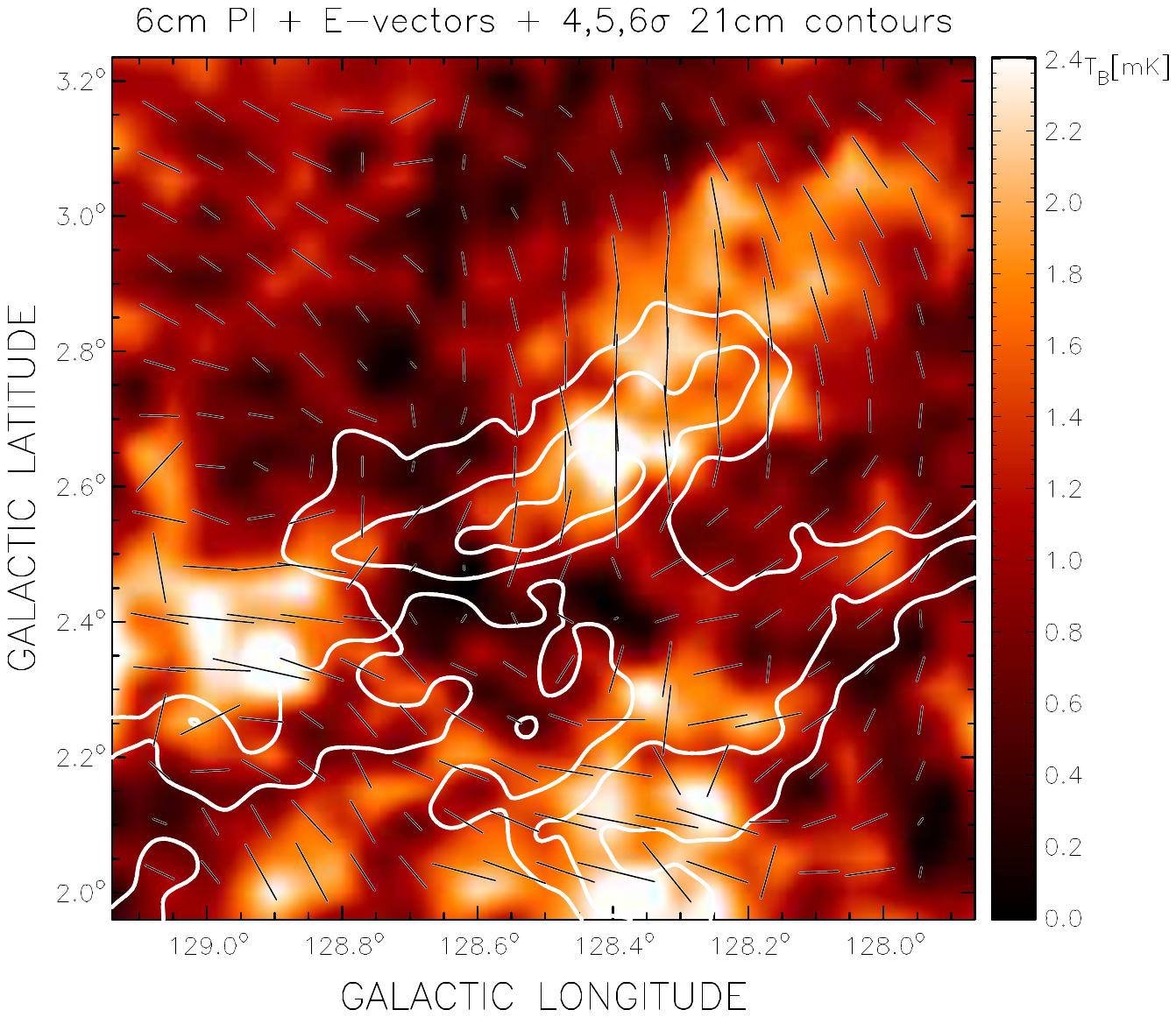}
\end{minipage}
\begin{minipage}{6.1cm}
\includegraphics[bb = 139 241 506 575,height=5.4cm]{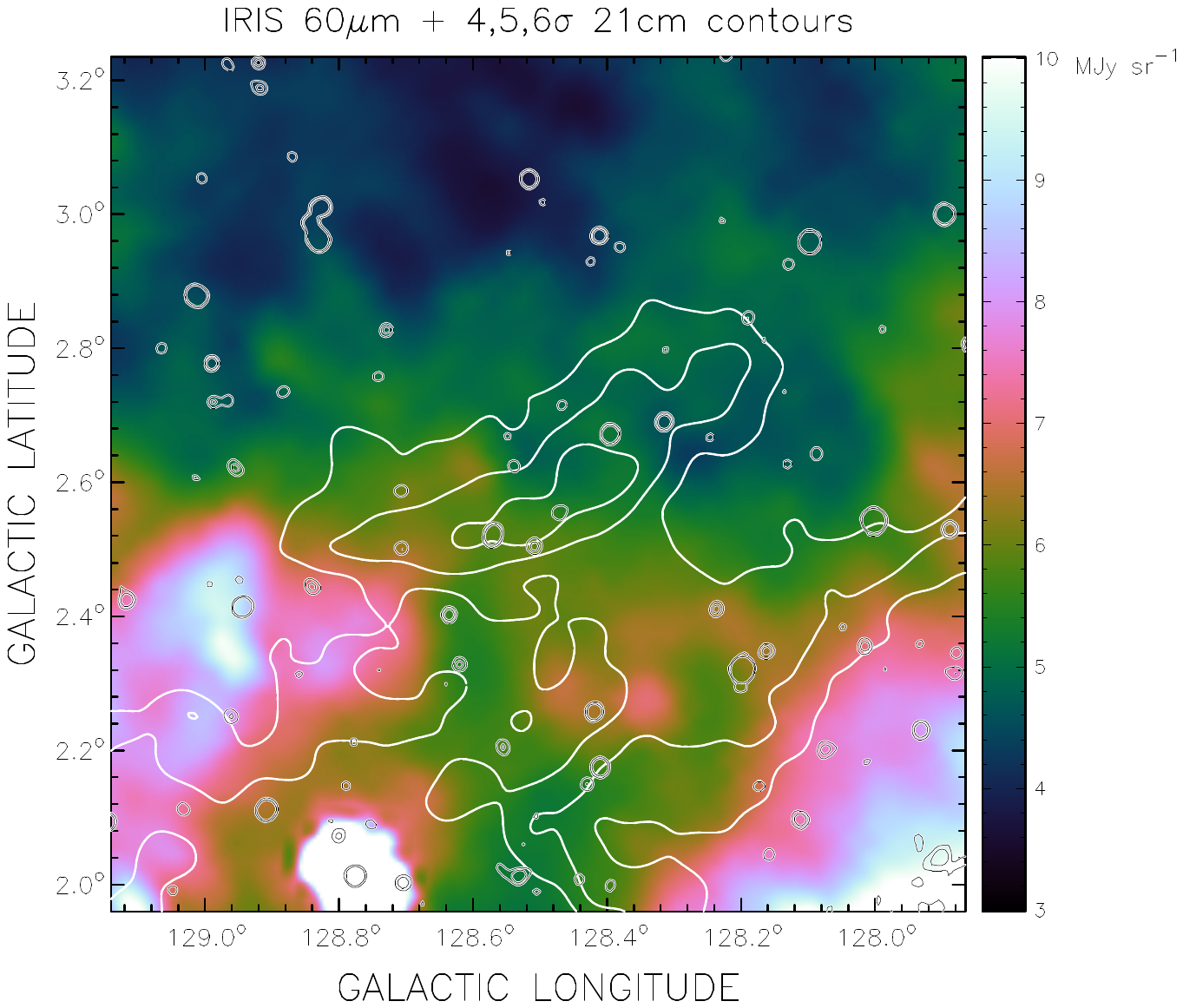}
\end{minipage}
}
\caption{\label{g128_maps} Total intensity maps of SNR candidate G128.5+2.6 at
327~MHz (top left), 408~MHz (top centre), 1420~MHz (top right), 2.7~GHz (middle 
left), and 4.8~GHz (middle centre). The polarized intensity map at 1420~MHz is 
shown in the middle row at right, and polarized intensity at 4.85~GHz is shown 
in the bottom row at left. Units of the radio maps are in Kelvin. An IRIS map 
of infrared emission at 60~$\mu$m (bottom right) covering the same region of 
the sky is shown for comparison. 
}
\end{figure*}

Total power maps at 4.8~GHz, 2.7~GHz, 1420~MHz, 408~MHz, and 327~MHz are shown 
in Figure~\ref{g128_maps}. In total power, G128.5$+$2.6 has a shell-like 
morphology with only one crescent-like limb detectable, appearing concave 
northwards. G128.5$+$2.6 is among the smaller of the five candidates, primarily
because there is no complete shell to be seen. The angular extent of the S 
shell limb is 40$\arcmin\times$22$\arcmin$ with a 30$\degr$ angle between the 
symmetry axis of the SNR and the Galactic Plane.

\begin{figure}[!ht]
\begin{center}
\includegraphics[scale=0.4,angle=0]{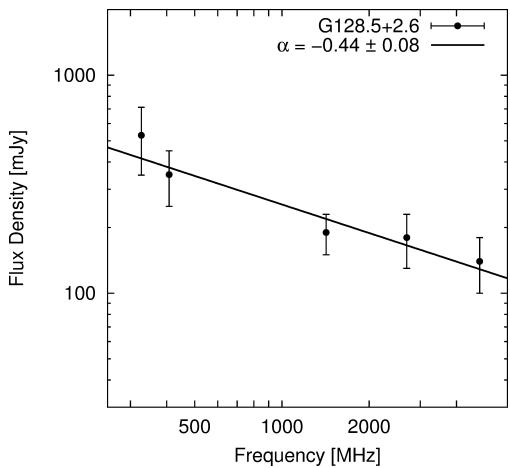}
\caption
{Integrated radio flux spectra for G128.5$+$2.6 using five frequencies and 
overlaid with power law fitted with weighted errors. 
} 
\label{g128flux}
\end{center}
\end{figure}

G128.5$+$2.6 has a moderately steep spectral index of 
$\alpha_{5}$~=~$-$0.44$\pm$0.08 (see Figure~\ref{g128flux}), typical of 
shell-type SNRs in the plane \citep{koth06}. The SNR is fairly well defined in 
all five radio frequencies that we use, appearing isolated in the direction
of minimal confusing foreground/background emission; therefore we deem our flux 
density measurements quite reliable. There is no optical or IR emission 
associated with the shell. Some isolated 21~cm polarized emission appears 
within the shell's Western side, and the \vec{B}-field vectors (perpendicular 
to the \vec{E}-field vectors shown in Figure~\ref{g128_maps}) with this 
emission run tangent to the shell and roughly parallel to the Galactic plane.
In the 9$\arcmin$-resolution $\lambda$6~cm map \citep[the complete data for
which are described by][]{xiao11}, a similar bright patch of PI 
appears in the shell's Western side; however this emission may or may not be 
related to the SNR since it appears connected to extended PI emission appearing 
outside the shell to the North and West. Because of the shell's small size, any 
emission belonging to it as shown in the $\lambda$6~cm polarization map in 
Figure~\ref{g128_maps} appears confused within the larger structure. Despite 
the inconclusive evidence of polarizaton in the $\lambda$6~cm map, the object's 
well-determined radio spectrum, its shell-like appearance with some moderately 
well-defined 21~cm PI emission, and an absence of IR emission make G128.5$+$2.6 
a strong candidate SNR.

\subsection{G149.5+3.2}
\begin{figure*}[!ht]
\centerline{
\begin{minipage}{6.3cm}
\includegraphics[bb = 121 241 506 575,height=5.4cm]{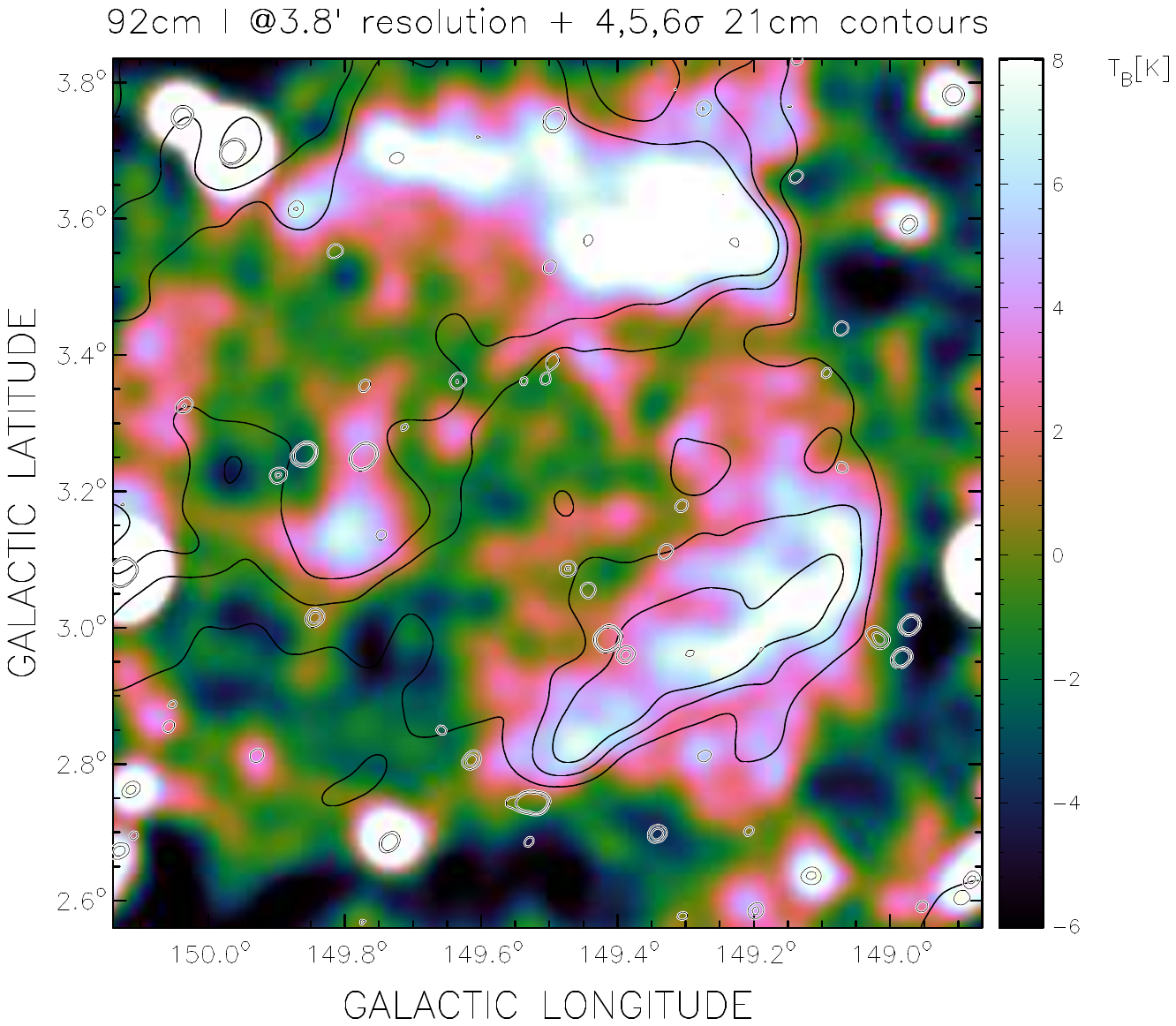}
\end{minipage}
\begin{minipage}{6.1cm}
\includegraphics[bb = 139 241 506 575,height=5.4cm]{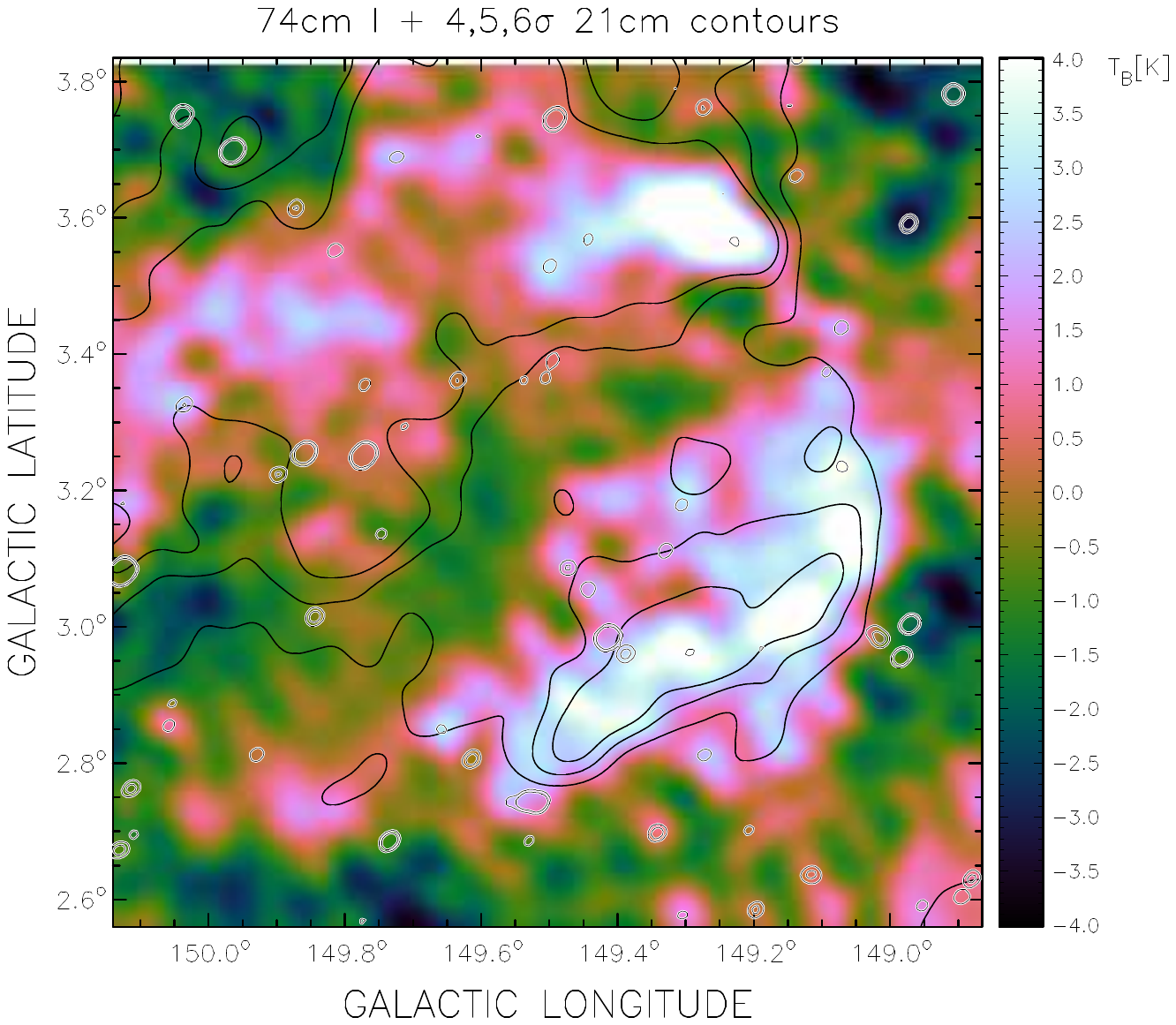}
\end{minipage}
\begin{minipage}{6.2cm}
\includegraphics[bb = 139 241 506 575,height=5.4cm]{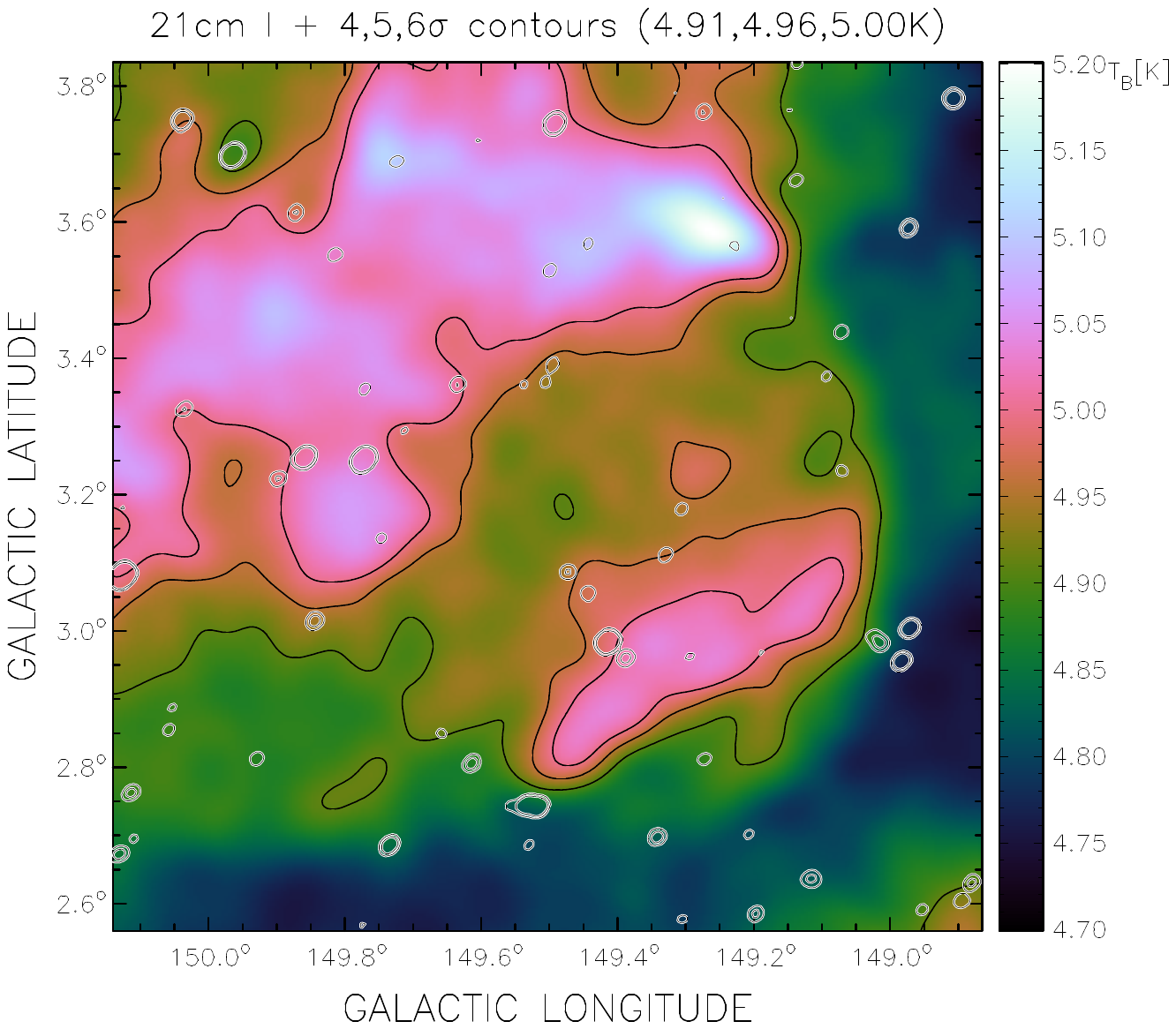}
\end{minipage}
}
\centerline{
\begin{minipage}{6.3cm}
\includegraphics[bb = 121 241 506 575,height=5.4cm]{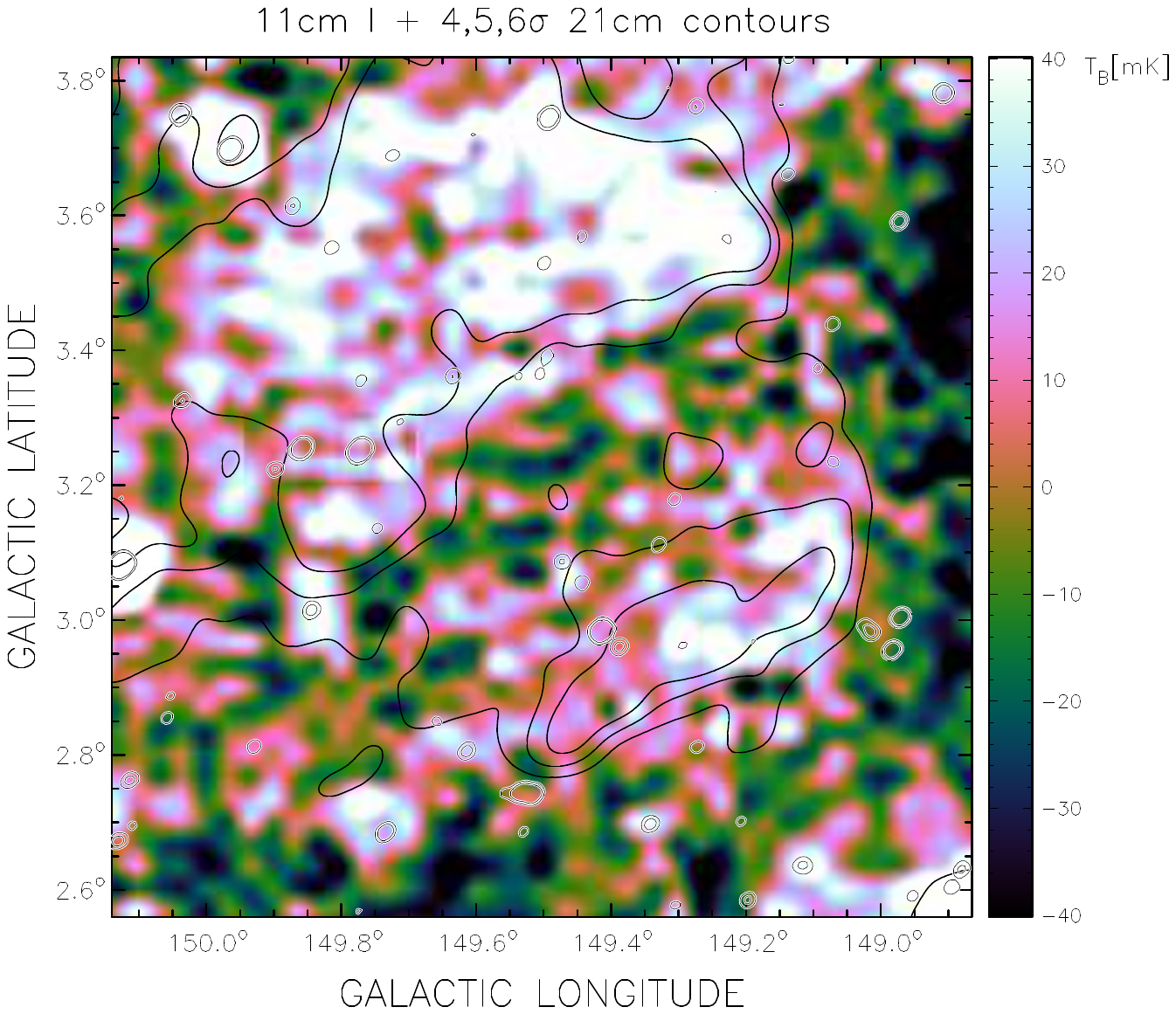}
\end{minipage}
\begin{minipage}{6.1cm}
\includegraphics[bb = 139 241 506 575,height=5.4cm]{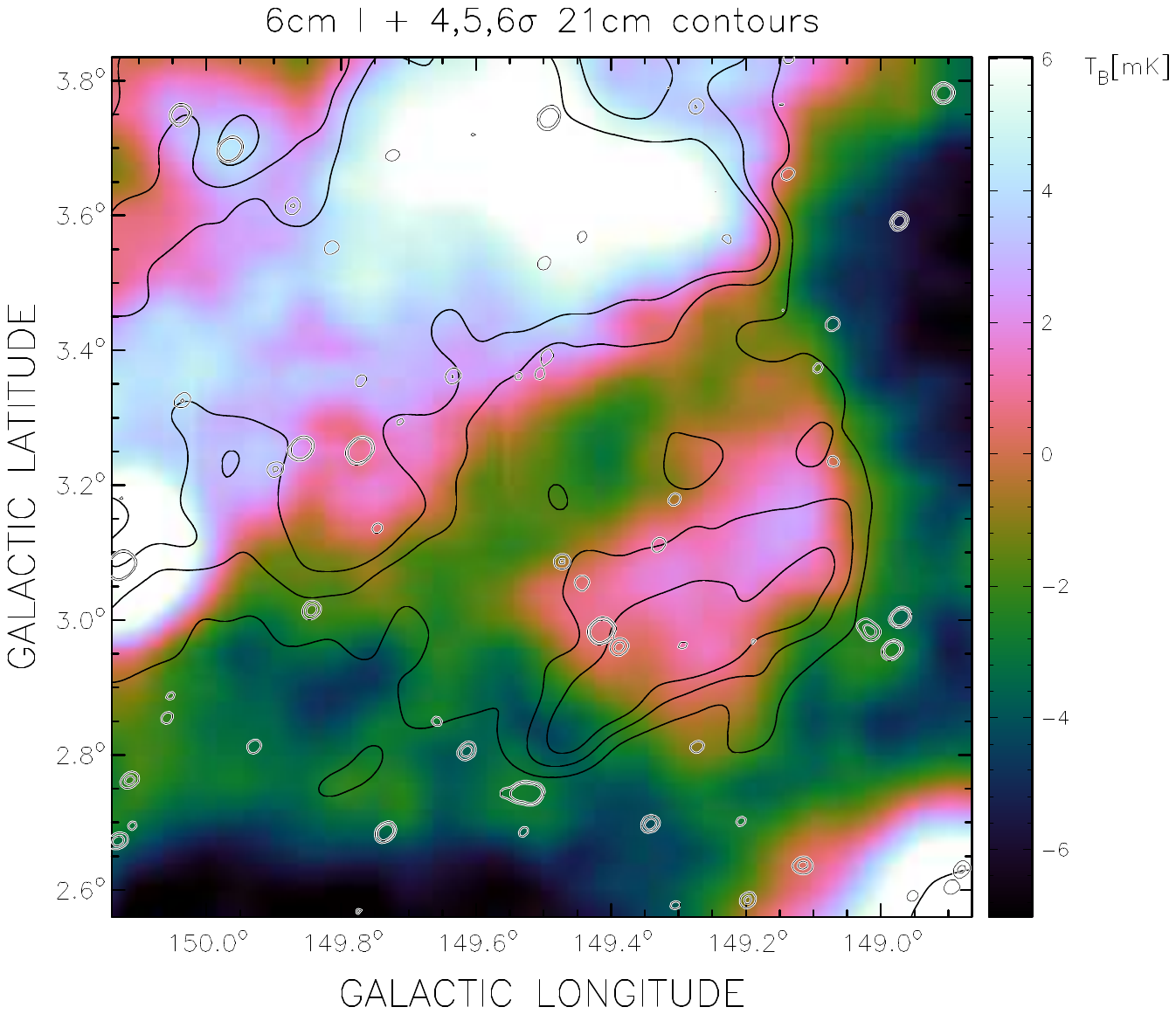}
\end{minipage}
\begin{minipage}{6.2cm}
\includegraphics[bb = 139 241 506 575,height=5.4cm]{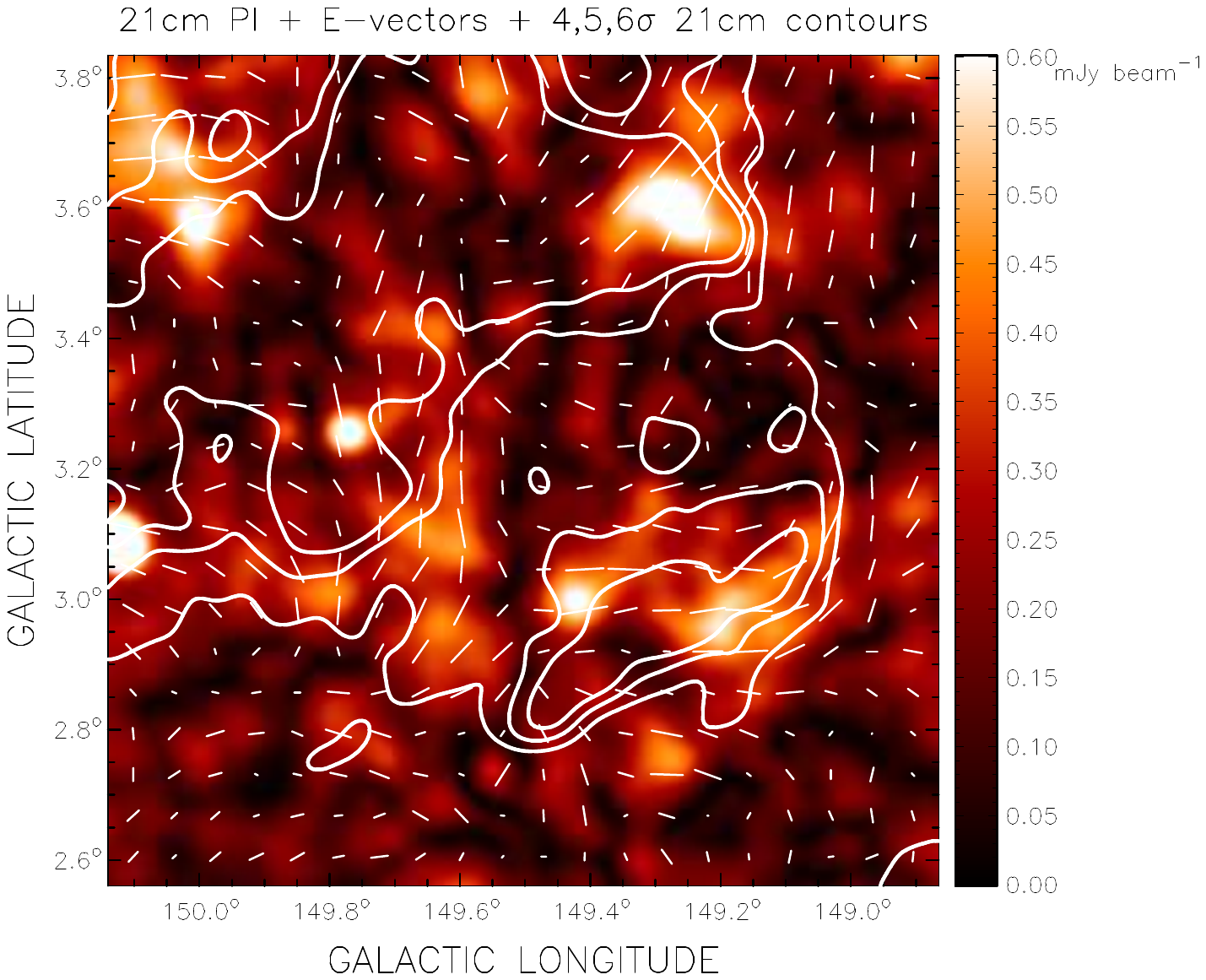}
\end{minipage}
}
\centerline{
\begin{minipage}{6.3cm}
\includegraphics[bb = 121 241 506 575,height=5.4cm]{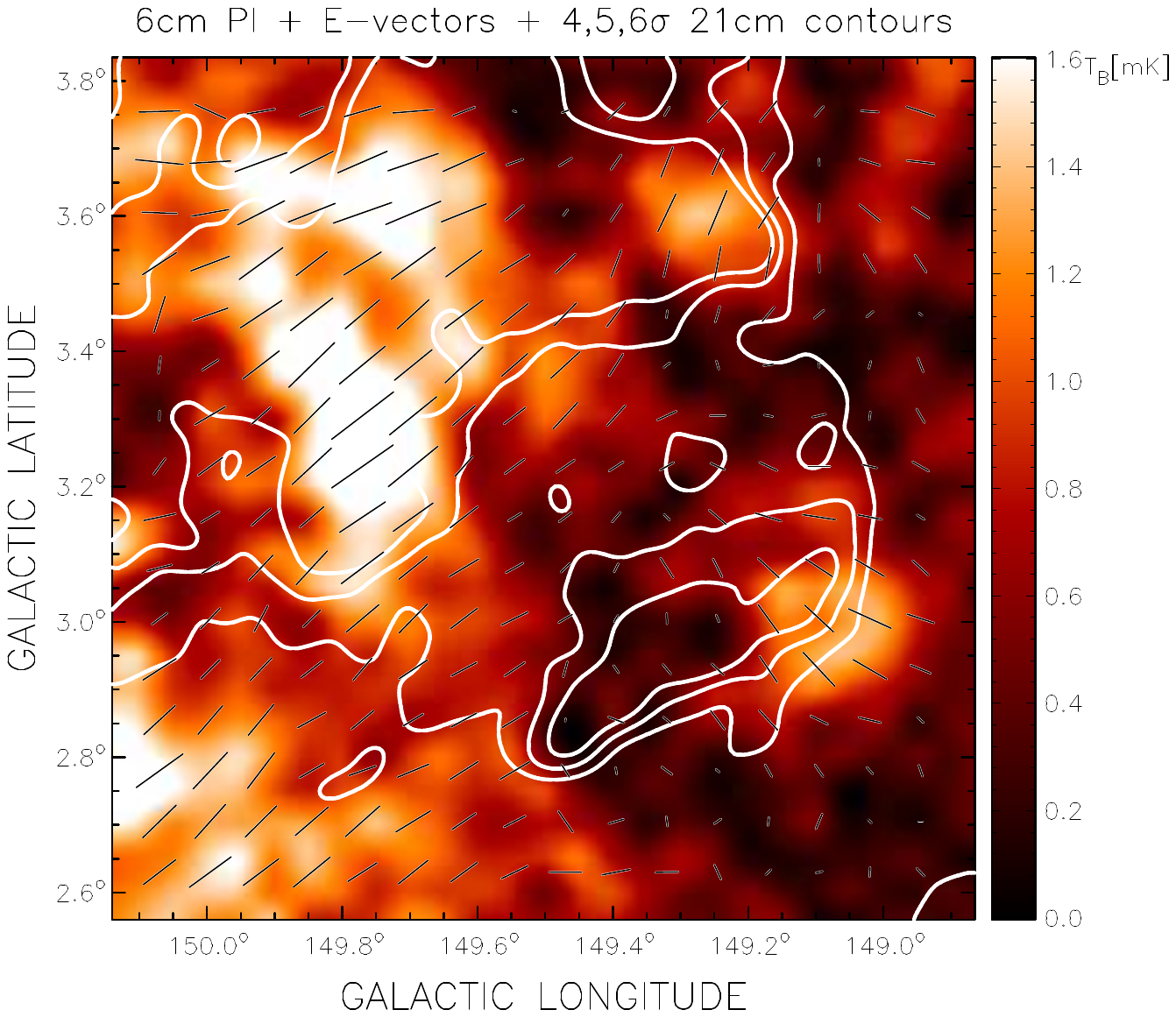}
\end{minipage}
\begin{minipage}{6.1cm}
\includegraphics[bb = 139 241 506 575,height=5.4cm]{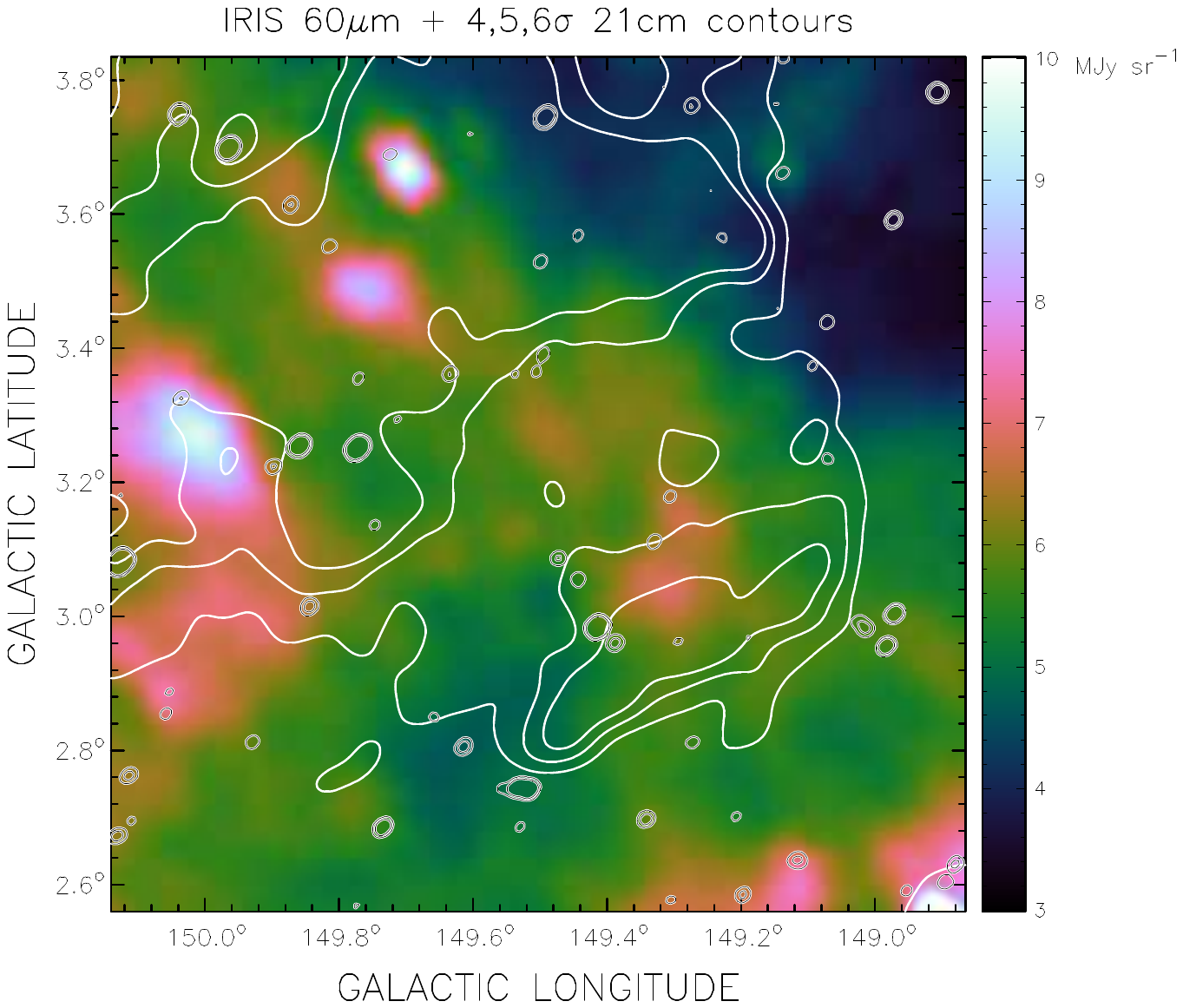}
\end{minipage}
}
\caption{\label{g149_maps} Same as in Figure~\ref{g128_maps} but for
SNR candidate G149.5+3.2. 
}
\end{figure*}

Total Power maps at 4.8~GHz, 2.7~GHz, 1420~MHz, 408~MHz, and 327~MHz, are shown 
in Figure~\ref{g149_maps}. In total power, G149.5$+$3.2 shows a bilateral 
structure with a N and S shell limb, the southern portion of which is better
defined and isolated from large-scale diffuse field emission. Among the five 
SNR candidates described in this paper, G149.5$+$3.2 is one of the larger in
angular size at 55$\farcm$6$\times$49$\farcm$3, with an angle between the 
symmetry axis of the SNR's Southern shell and Galactic plane of roughly 
40$\degr$.

\begin{figure}[!ht]
\begin{center}
\includegraphics[scale=0.4,angle=0]{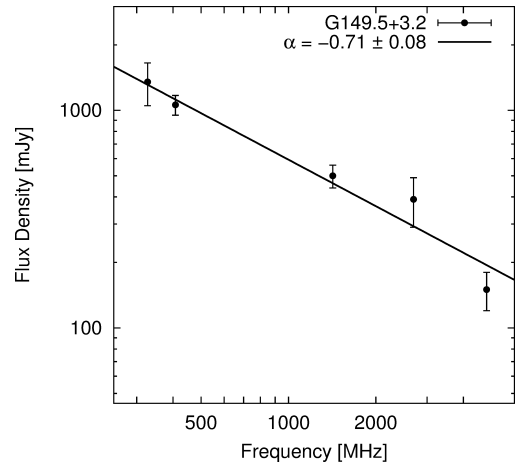}
\caption
{Same as Figure~\ref{g128flux} but for G149.5$+$3.2.} 
\label{g149flux}
\end{center}
\end{figure}

All total power maps show emission well correlated to the 21~cm contours of the 
object's limb. The integration area chosen to determine the spectral index 
included the brightened S limb and the well-delineated central round portion of 
the object, but we exclude the bright emission in the N edge where the extent
of the shell is ambiguous and possibly confused with extended diffuse Galactic
emission. The flux measurements taken included variations on the size and shape 
of the area to quantify the uncertainty in the flux due to the ambiguity in
extent of emission. The spectral index is steep: 
$\alpha_{5}$~=~$-$0.71$\pm$0.08 (see Figure~\ref{g149flux}), typical of a young 
SNR. 

There is isolated 21~cm PI emission appearing confined within the well-defined 
S shell half with none outside, having \vec{E} vectors parallel with the Plane 
and therefore \vec{B} vectors almost radial with respect to the shell. This 
21~cm map is to be compared with the 6~cm PI map \citep[the complete
data for which is described by][]{gao10}, which also indicates PI 
emission within the S shell's boundary. the 6~cm PI emission is somewhat 
correlated with the 21~cm PI, but only roughly due to the lower 
9$\arcmin$-resolution. The $\lambda$6~cm PI emission shows projected \vec{B} 
vectors parallel to the shell (i.e. a tangentially oriented field) and nearly 
parallel to the Plane. Clearly this emission has suffered more rotation at 
$\lambda$21~cm by the intervening foreground.
There is no IR emission in and around G149.5$+$3.2, and no apparent optical
(red) emission. The well determined radio spectrum, its limb-brightened
appearance in several frequencies, and the presence of
spatially correlated PI at both 6 and 21~cm wavelengths make this a strong
SNR candidate. 

\subsection{G150.8+3.8}
\begin{figure}[!ht]
\begin{center}
\includegraphics[scale=0.4,angle=0]{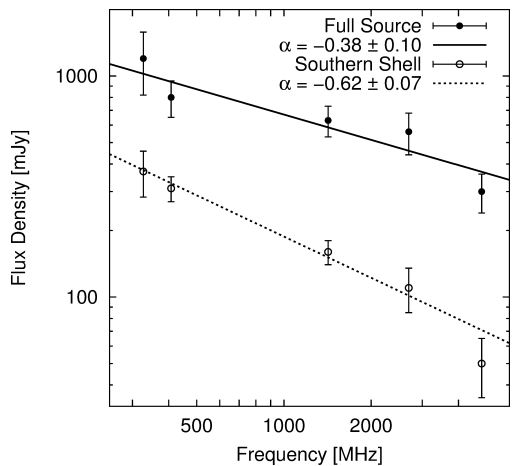}
\caption
{Same as Figure~\ref{g128flux} but for G150.8$+$3.8; both the full source and 
the Southern shell only.
} 
\label{g150flux}
\end{center}
\end{figure}

\begin{figure*}[!ht]
\centerline{
\begin{minipage}{6.3cm}
\includegraphics[bb = 121 241 506 575,height=5.4cm]{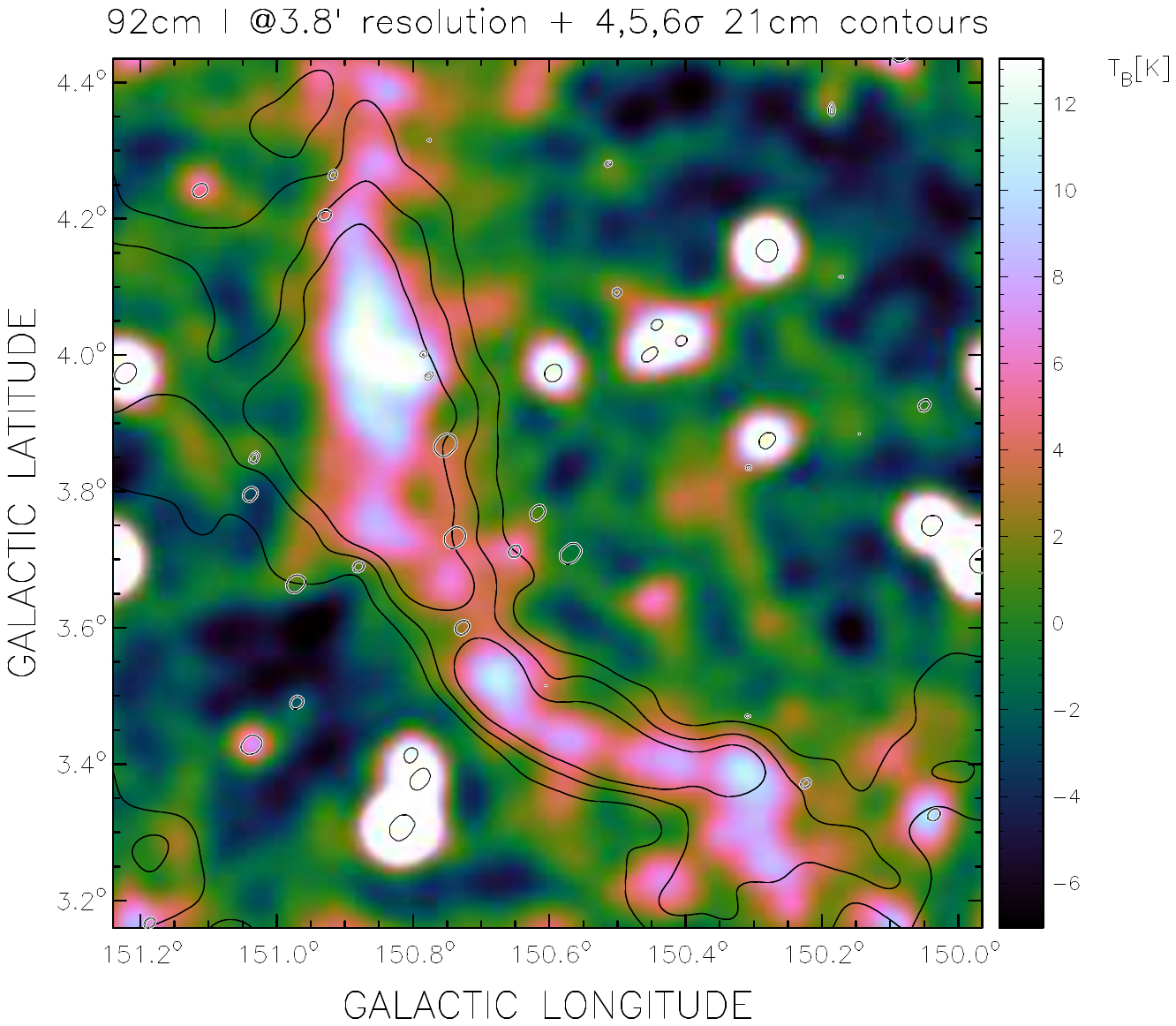}
\end{minipage}
\begin{minipage}{6.1cm}
\includegraphics[bb = 139 241 506 575,height=5.4cm]{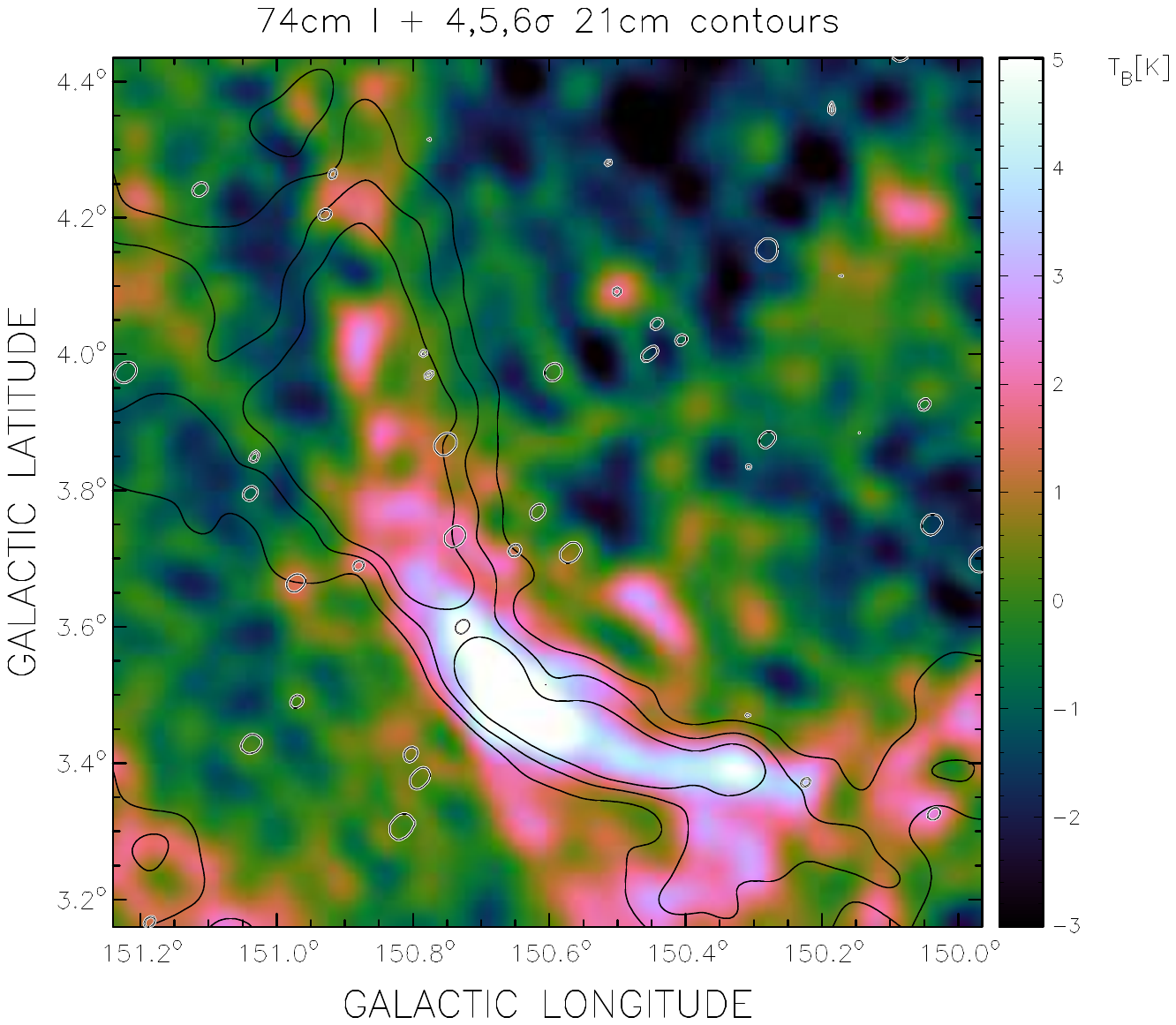}
\end{minipage}
\begin{minipage}{6.2cm}
\includegraphics[bb = 139 241 506 575,height=5.4cm]{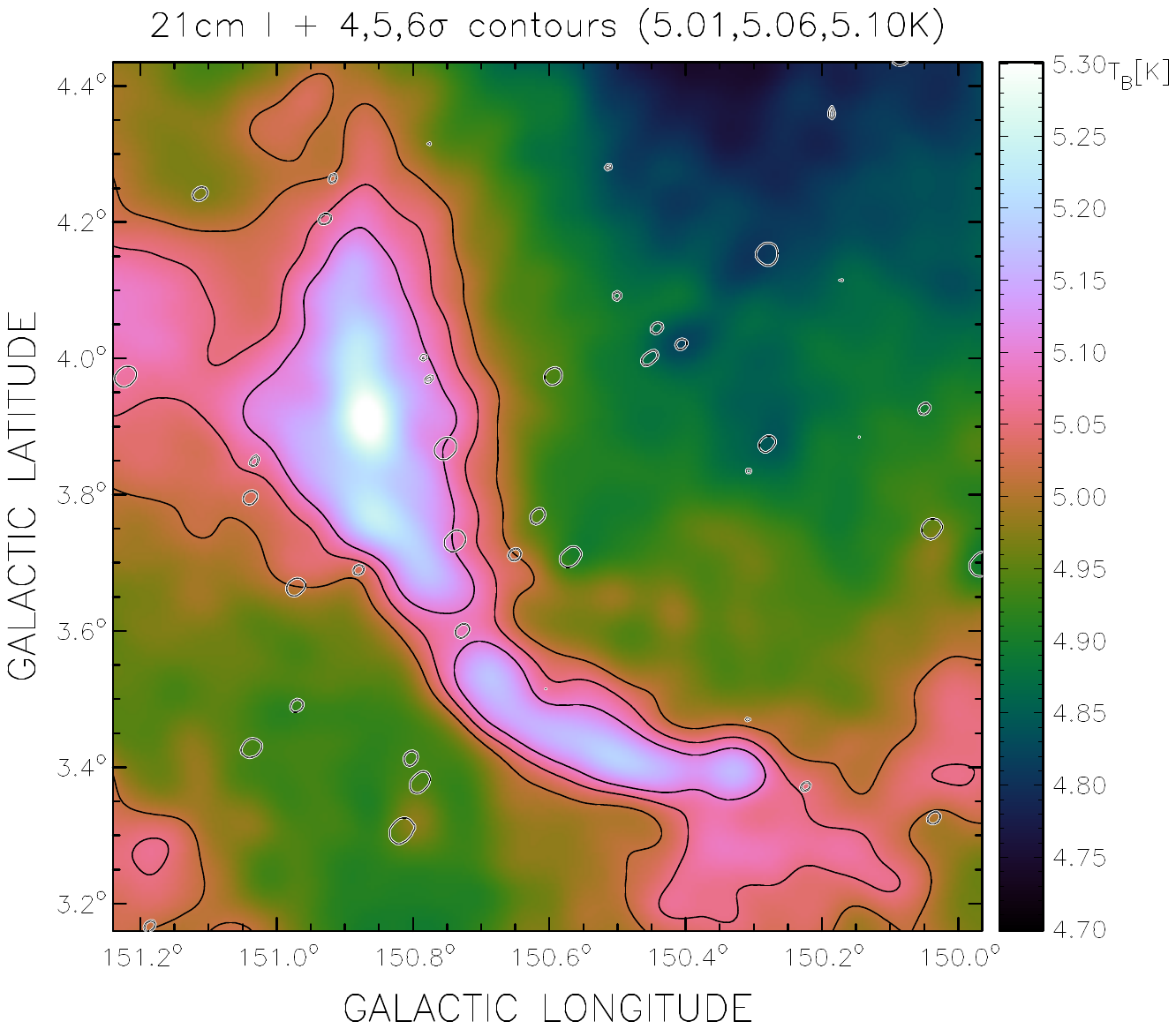}
\end{minipage}
}
\centerline{
\begin{minipage}{6.3cm}
\includegraphics[bb = 121 241 506 575,height=5.4cm]{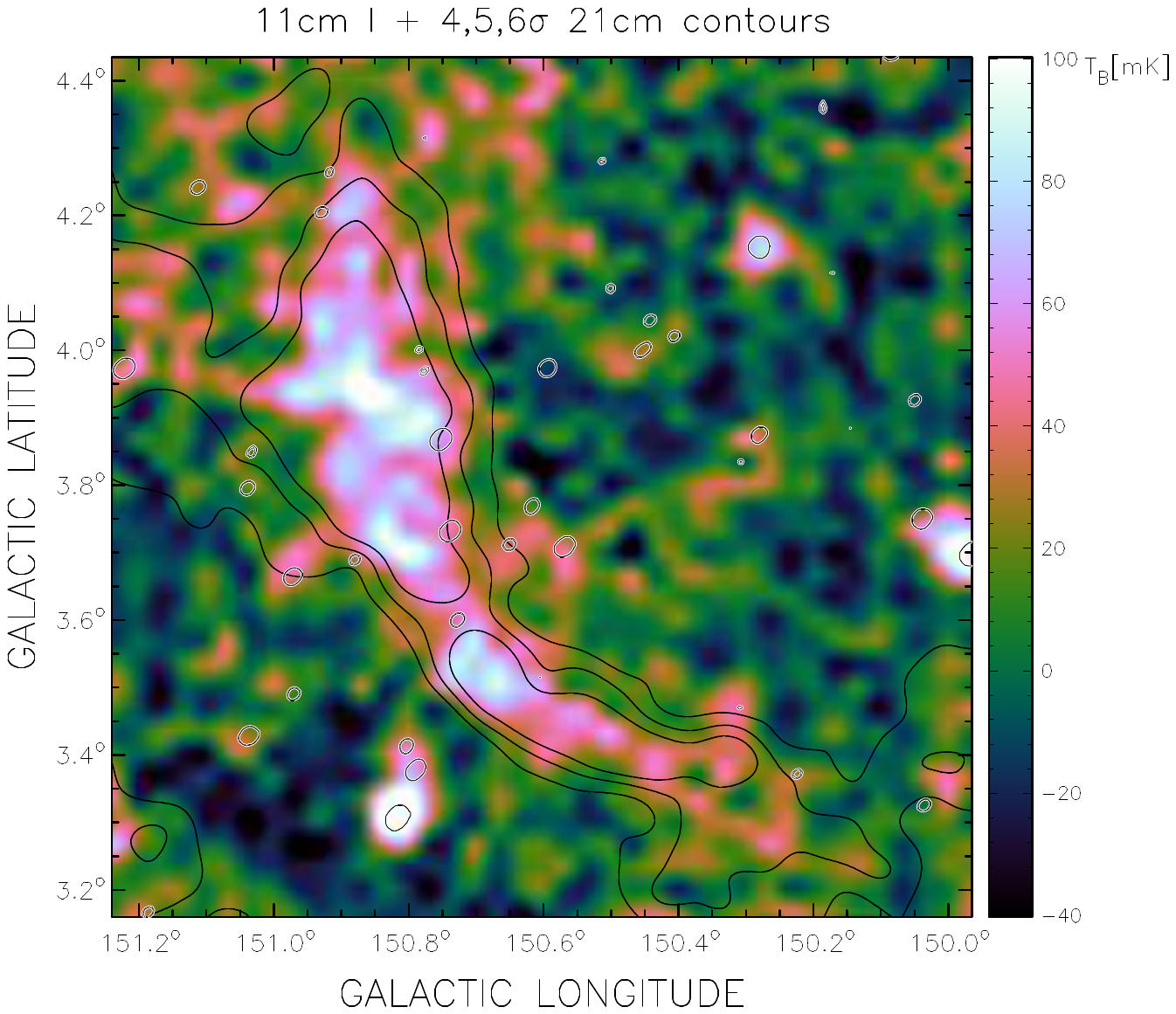}
\end{minipage}
\begin{minipage}{6.1cm}
\includegraphics[bb = 139 241 506 575,height=5.4cm]{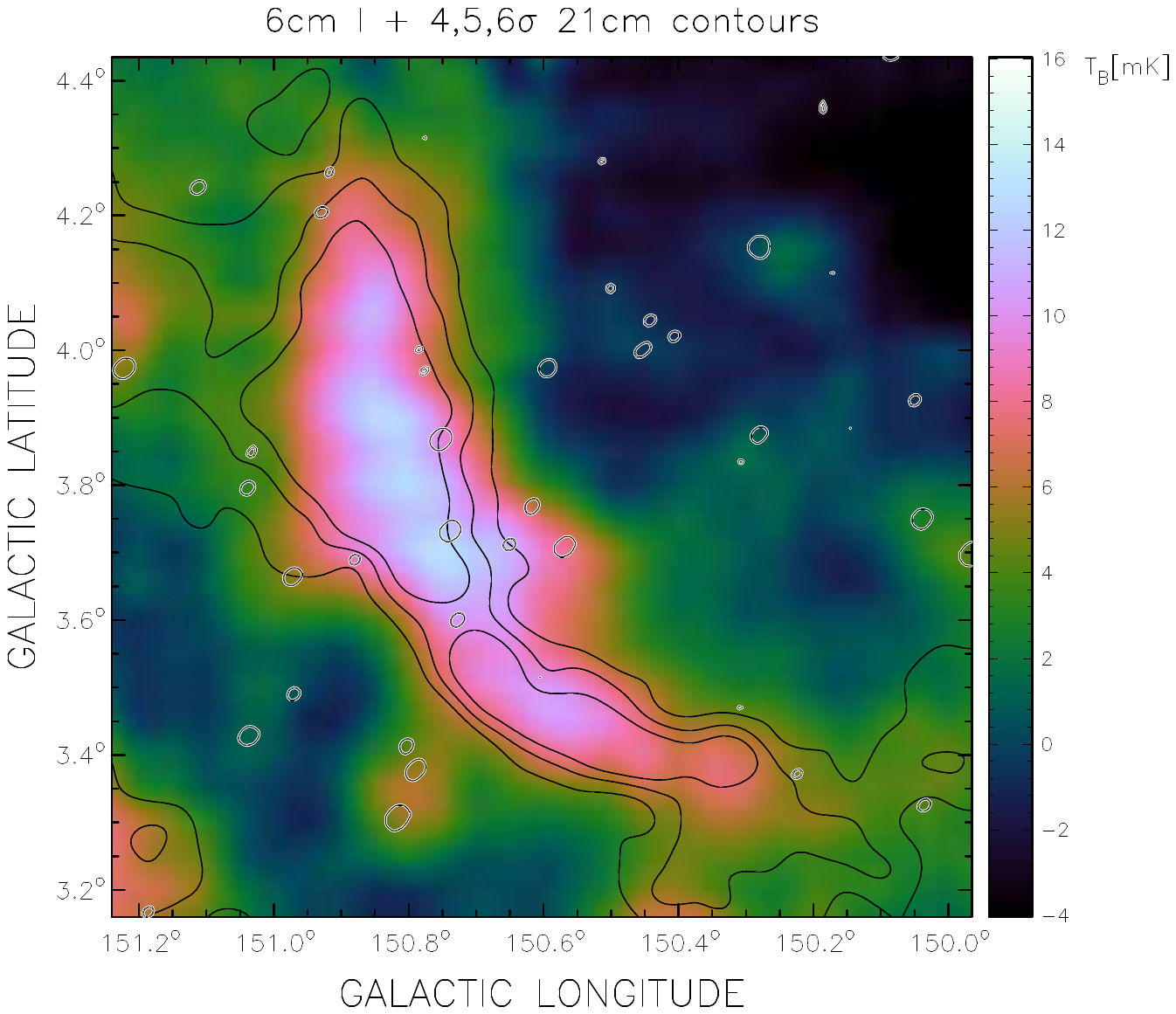}
\end{minipage}
\begin{minipage}{6.2cm}
\includegraphics[bb = 139 241 506 575,height=5.4cm]{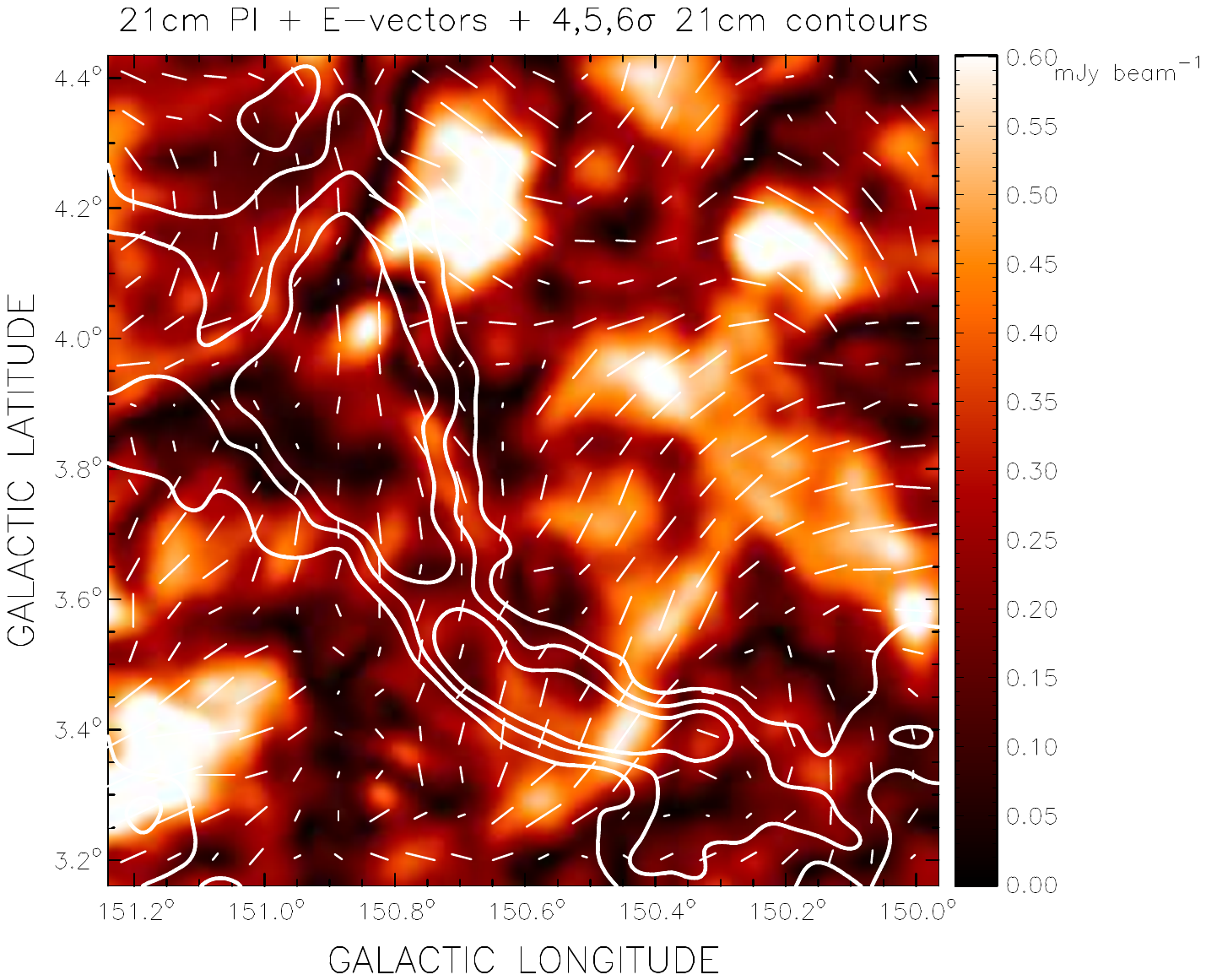}
\end{minipage}
}
\centerline{
\begin{minipage}{6.3cm}
\includegraphics[bb = 121 241 506 575,height=5.4cm]{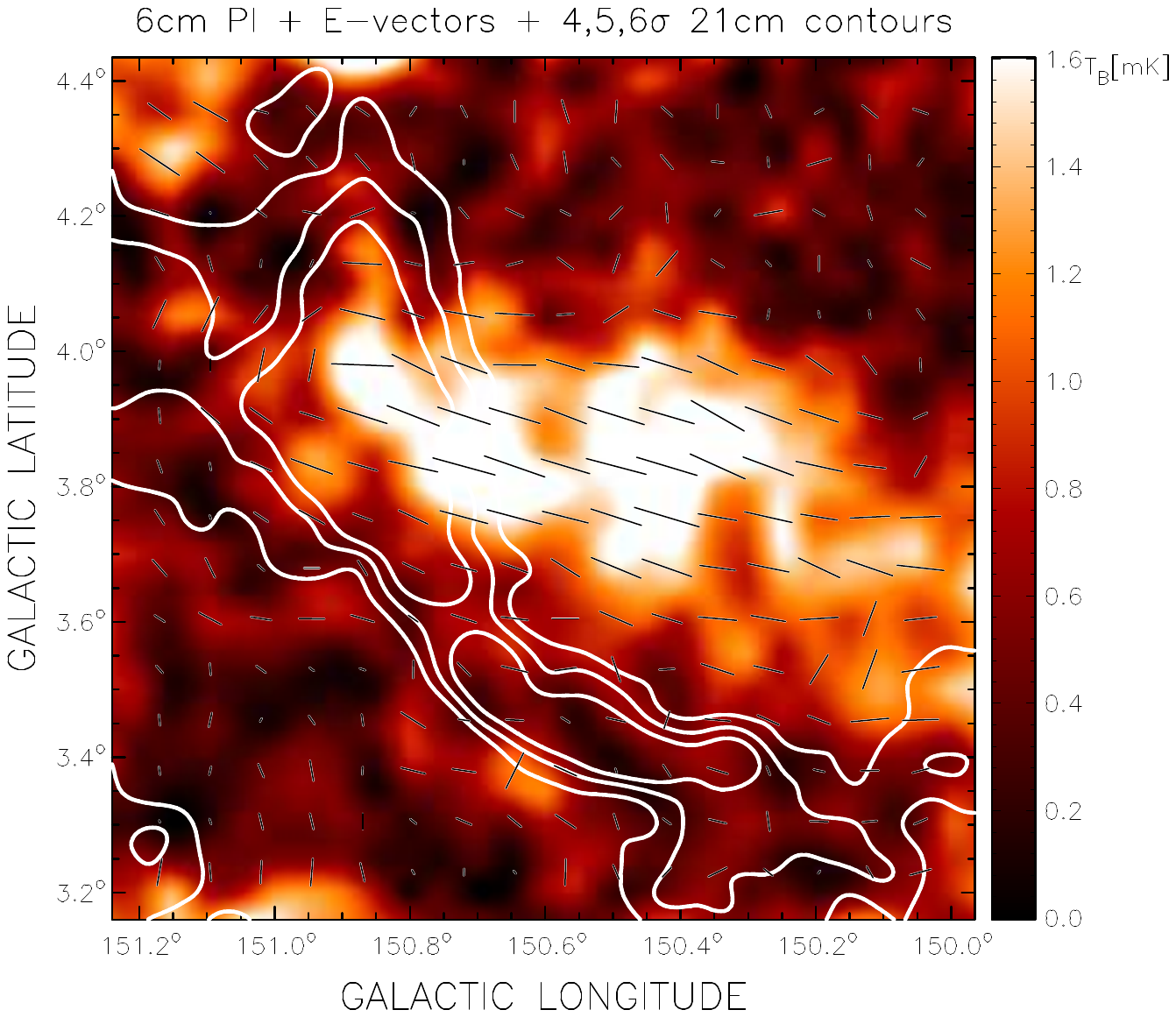}
\end{minipage}
\begin{minipage}{6.1cm}
\includegraphics[bb = 139 241 506 575,height=5.4cm]{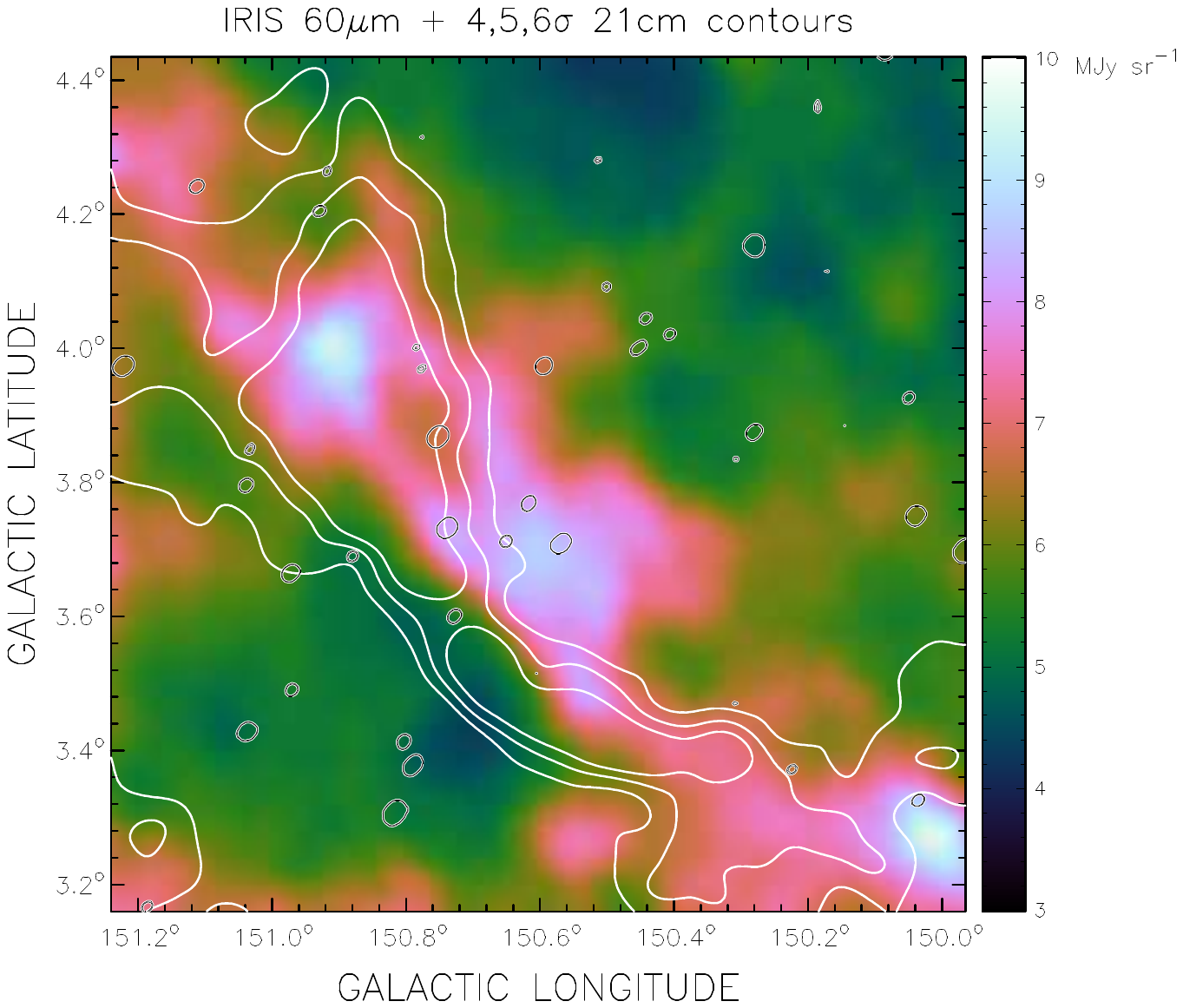}
\end{minipage}
\begin{minipage}{6.1cm}
\includegraphics[bb = 139 241 506 575,height=5.4cm]{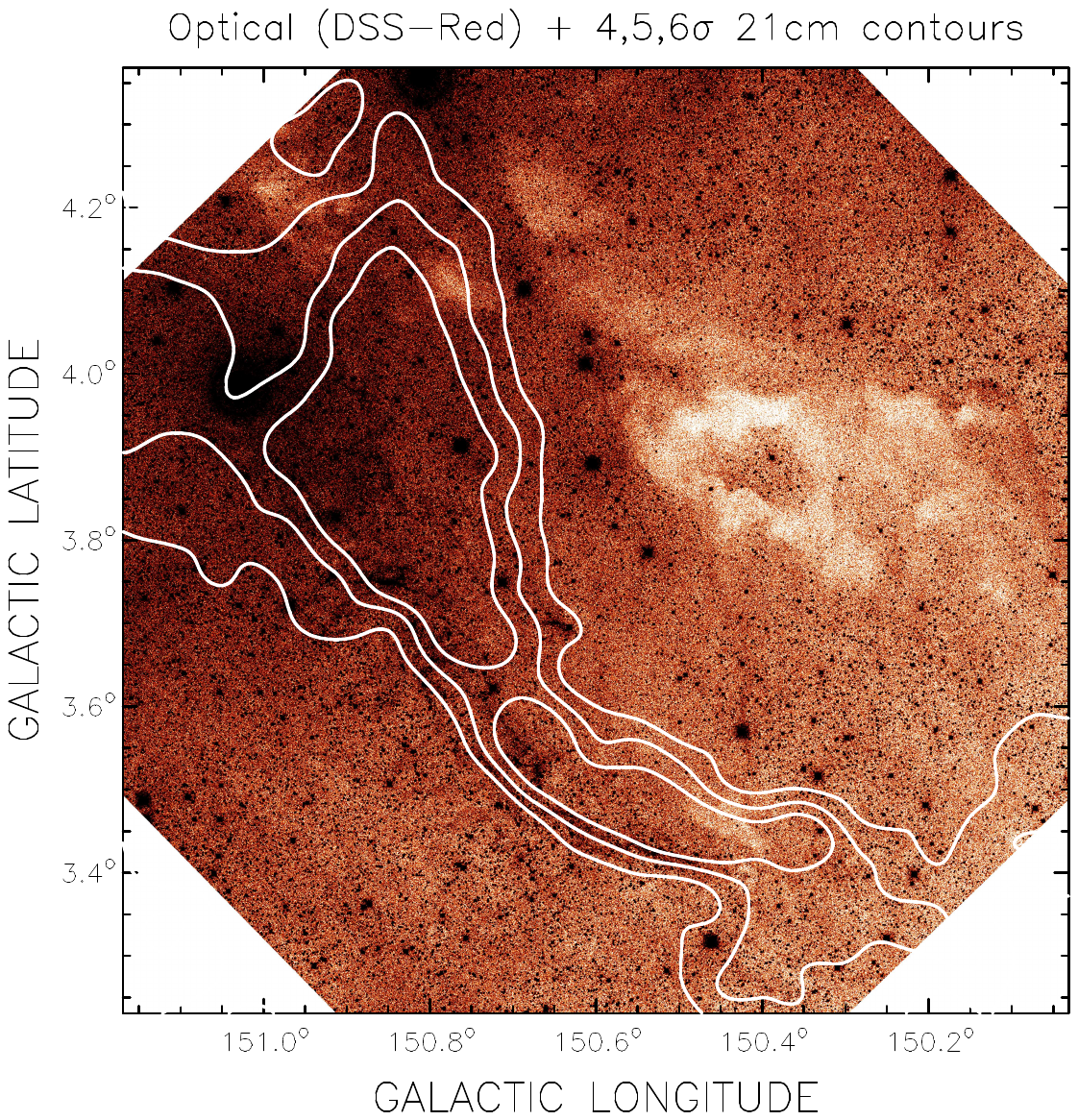}
\end{minipage}
}
\caption{\label{g150_maps} Same as in Figure~\ref{g128_maps} but for
SNR candidate G150.8+3.8. An IRIS map of infrared 
emission at 60~$\mu$m (bottom centre) and a DSS2 Red map (bottom right) 
covering the same region of the sky are shown for comparison. 
}
\end{figure*}

Total Power maps at 4.8~GHz, 2.7~GHz, 1420~MHz, 408~MHz, and 327~MHz, are shown in 
Figure~\ref{g150_maps}. In 1420~MHz total power, G150.8$+$3.8 appears as two
tenuously connected crescents of emission: one in the North (centre 
$\ell=$150.9$\degr$, $b=+$3.9$\degr$) elongated 90$\degr$ to the plane and one 
in the South (centre $\ell=$150.5$\degr$, $b=+$3.4$\degr$) elongated 
more-or-less along the Plane. It is unclear if these crescents form part of the 
same Eastern shell-edge of a single large object, or are two separate adjacent 
objects. In the DSS optical map (see Fig.~\ref{g150_maps}) the S shell encloses 
strikingly thin red filaments stretching parallel to the radio shell, while in 
the N shell similar red filaments are also seen within the shell; however, N
shell filaments appear atop some very diffuse red nebulosity that extends 
beyond the shell's radio contours and does not seem to be correlated. Both 
shells seem to show possible post-shock optical emission from an SNR shock, 
with the N shell strands seen towards unrelated thermal emission along the same 
line-of-sight. The 60~$\mu$m infrared map (see Figure~\ref{g150_maps}) 
indicates some emission towards the most Northern portion of the 1420~MHz 
contours, although the overall IR emission does not appear related to the 
shell-shaped contours. Indeed, taken together the overall integrated spectral 
index for G150.8$+$3.8 is the least steep of all five candidates, at 
$\alpha_{5}$~=~$-$0.38$\pm$0.10 (see Figure~\ref{g150flux}), but the thin S 
shell integrated alone shows a steeper spectrum at $\alpha_{5}=-$0.62$\pm$0.07. 
Polarized emission in the $\lambda$6~cm map appears to be associated with the 
large-scale foreground/background, as does the polarized emission in the 
1420~MHz map (see maps Figure~\ref{g150_maps}). Therefore, we cannot conclude 
that any of the polarized emission belongs to the observed shells. If the 
large-scale thermal emission observed in the optical and IR maps is in the 
foreground, it may be depolarizing any ordered PI emission from the shell.  

The presence of thin filamentary strands of red optical emission combined with 
the semi-circular shell-like appearance and distinctly non-thermal radio 
spectrum make G150.8$+$3.8 a strong SNR candidate.


\subsection{G160.1$-$1.1}
\begin{figure}[!ht]
\begin{center}
\includegraphics[scale=0.4,angle=0]{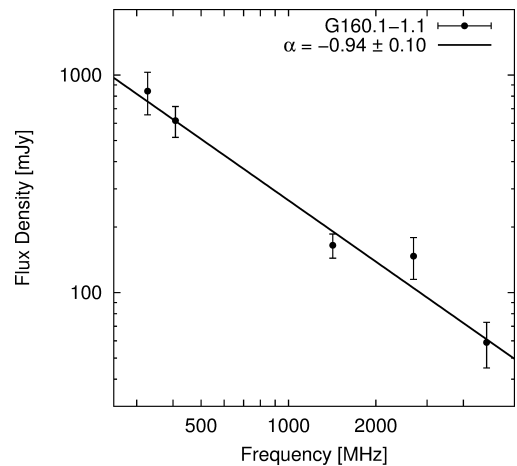}
\caption
{Same as Figure~\ref{g128flux} but for G160.1$-$1.1. 
} 
\label{g160flux}
\end{center}
\end{figure}

\begin{figure*}[!ht]
\centerline{
\includegraphics[bb = 121 241 506 575,height=5.4cm]{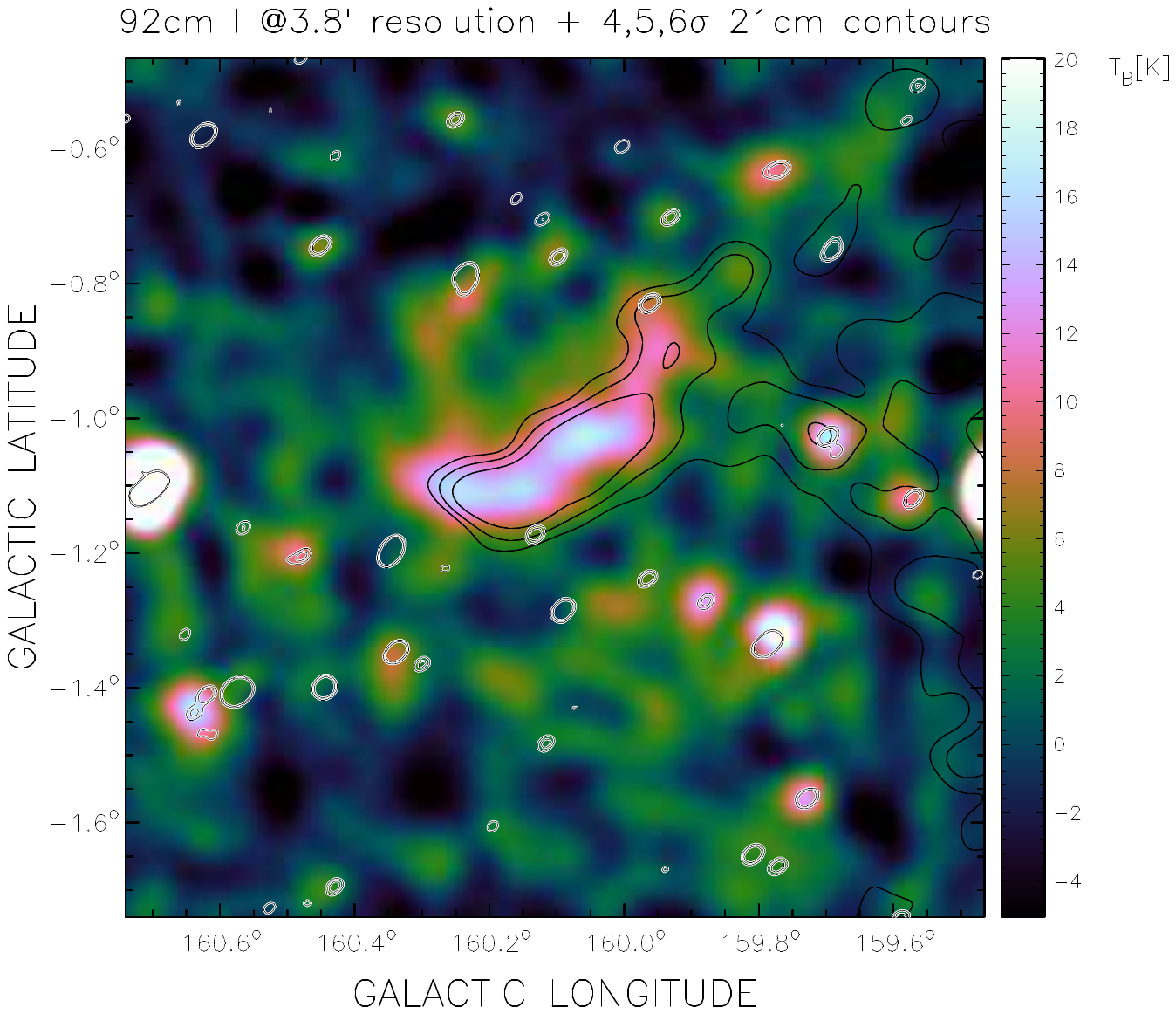}
\includegraphics[bb = 139 241 506 575,height=5.4cm]{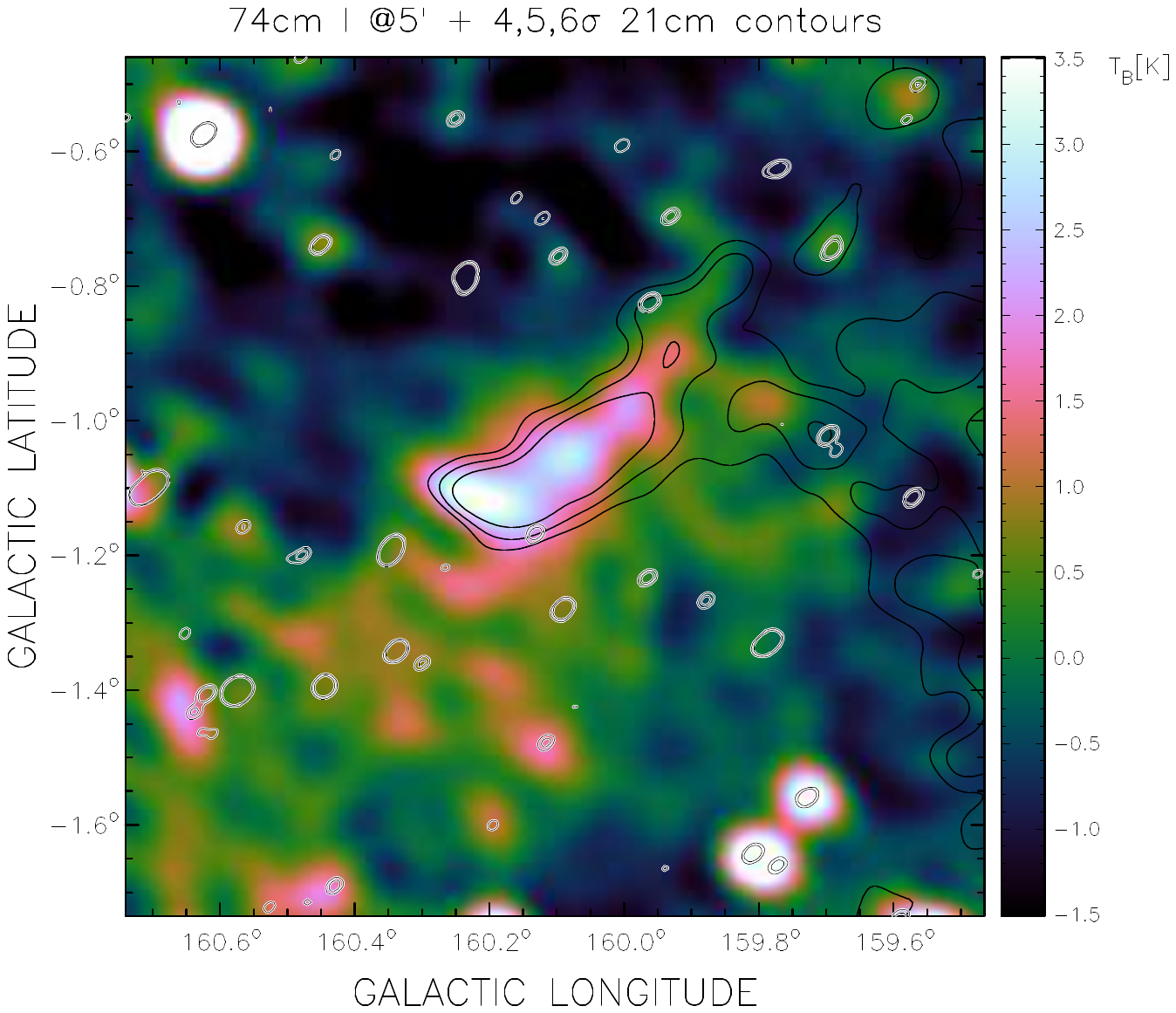}
\includegraphics[bb = 139 241 506 575,height=5.4cm]{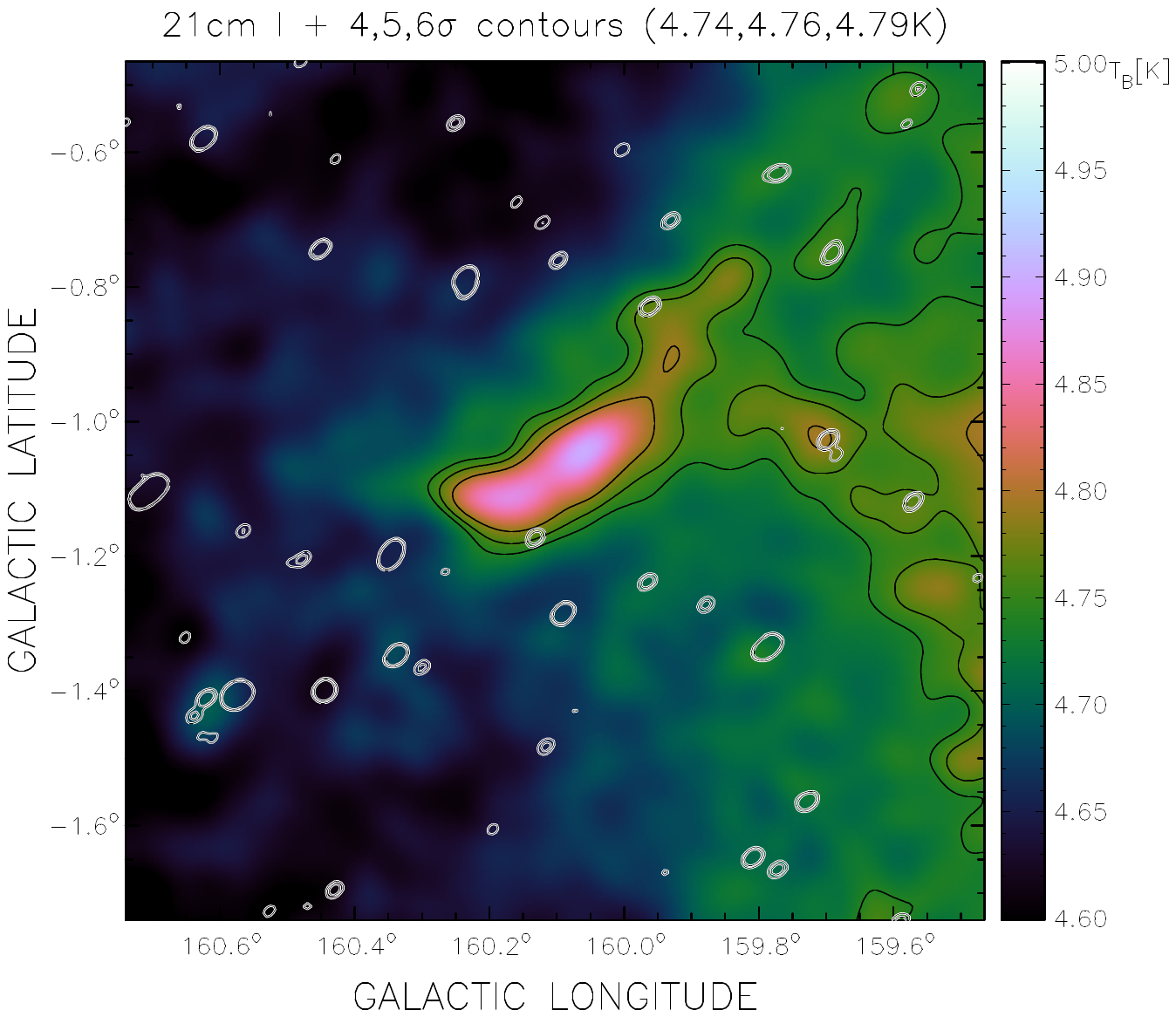}
}
\centerline{
\includegraphics[bb = 121 241 506 575,height=5.4cm]{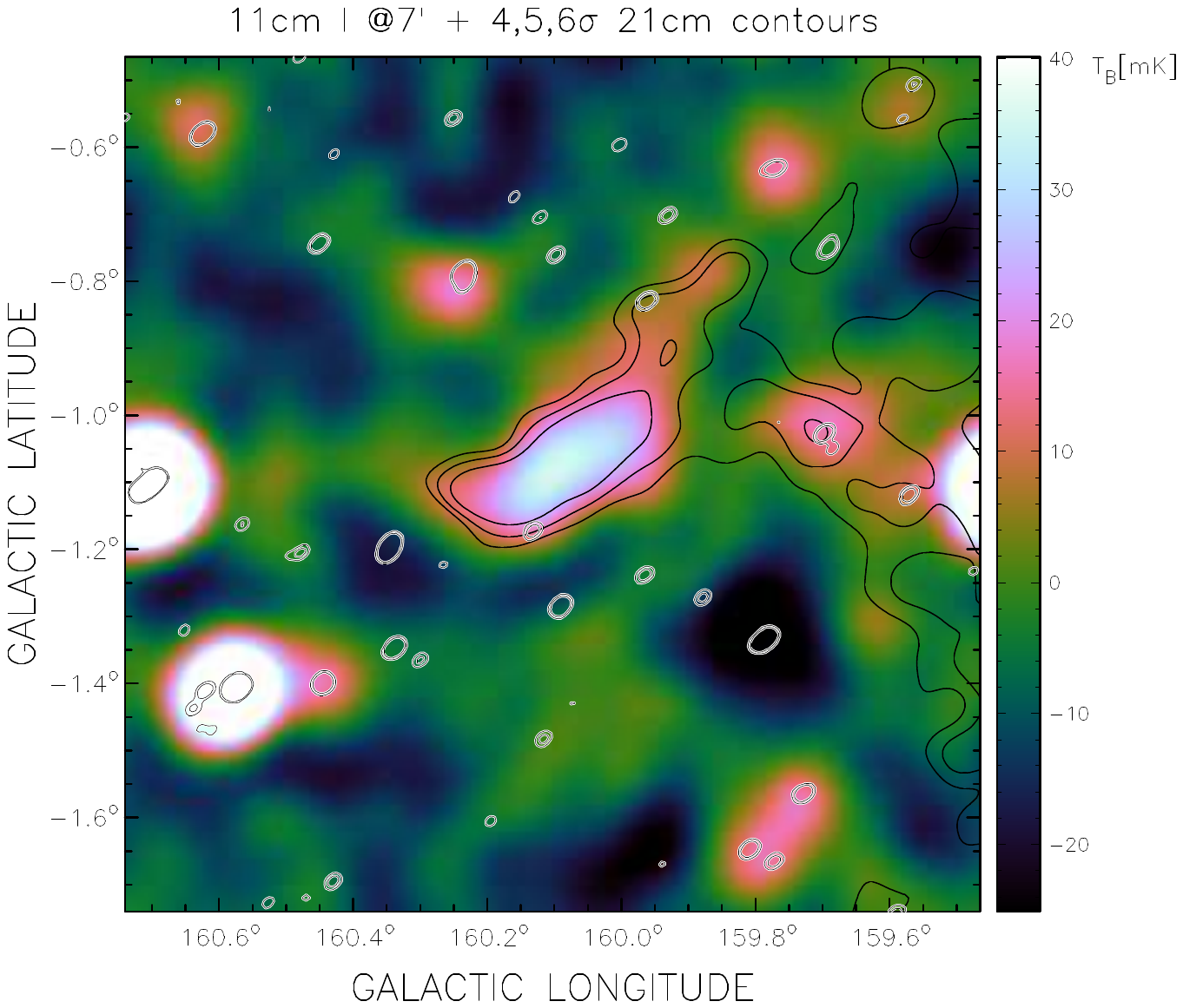}
\includegraphics[bb = 139 241 506 575,height=5.4cm]{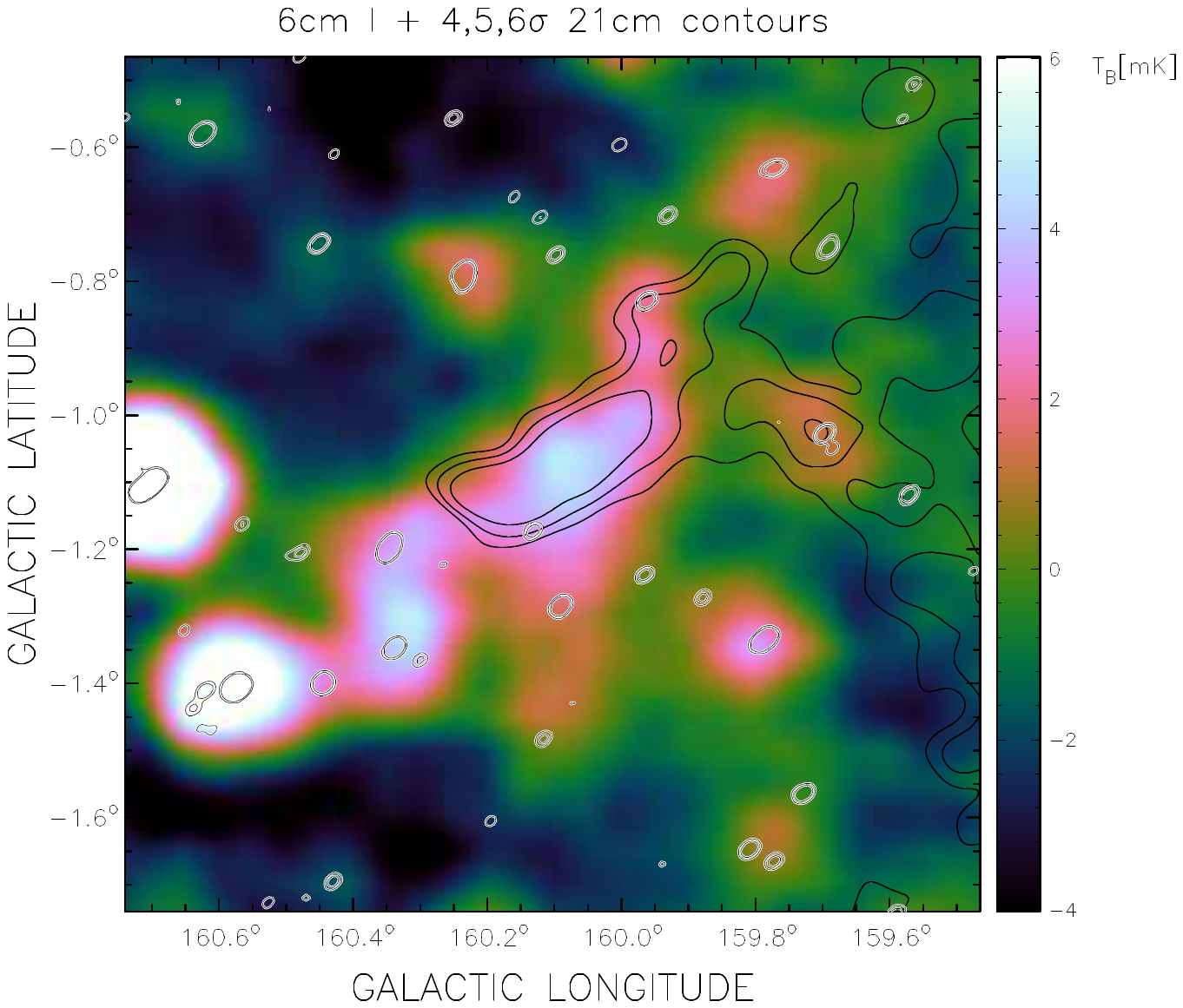}
\includegraphics[bb = 139 241 506 575,height=5.4cm]{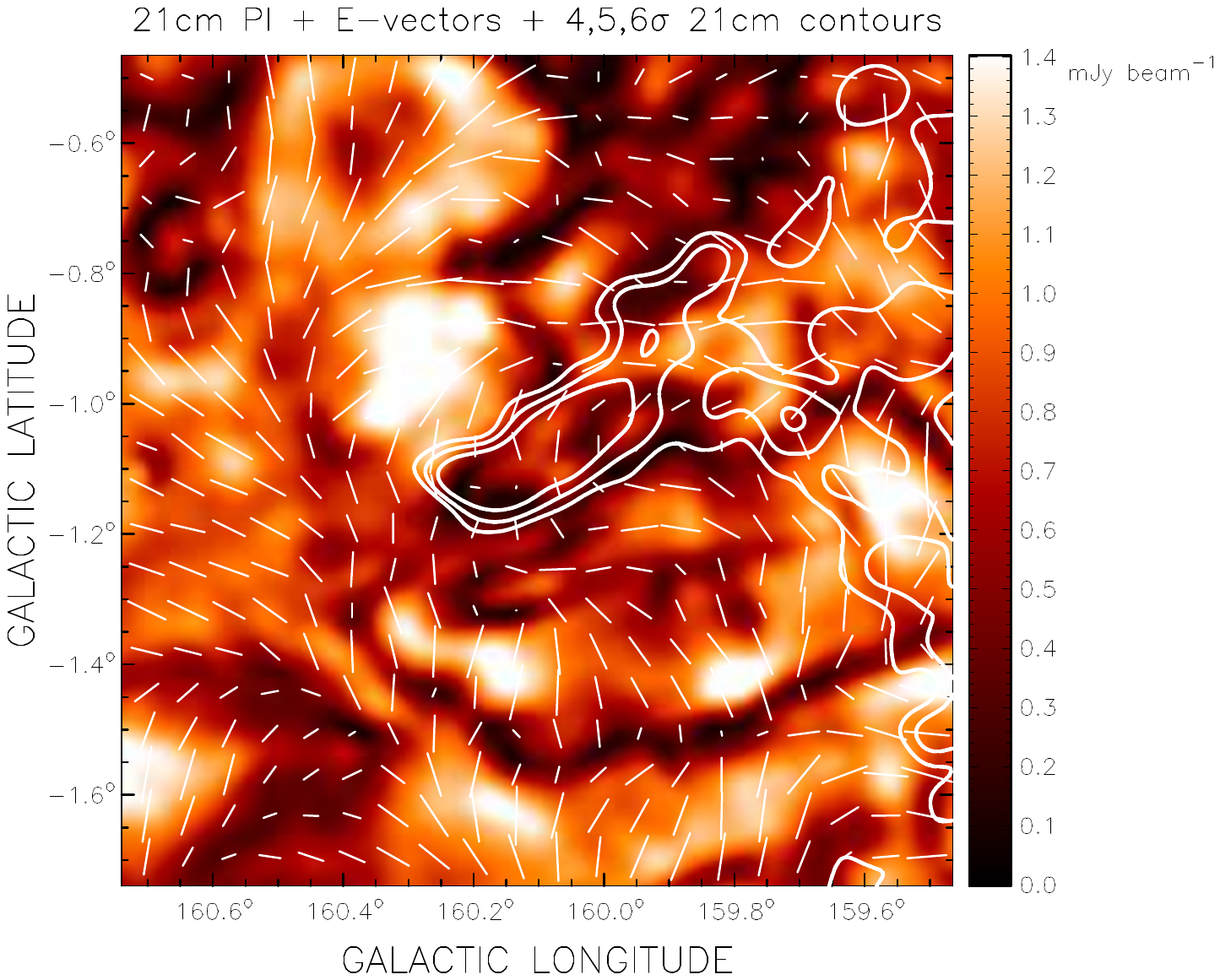}
}
\centerline{
\includegraphics[bb = 121 241 506 575,height=5.4cm]{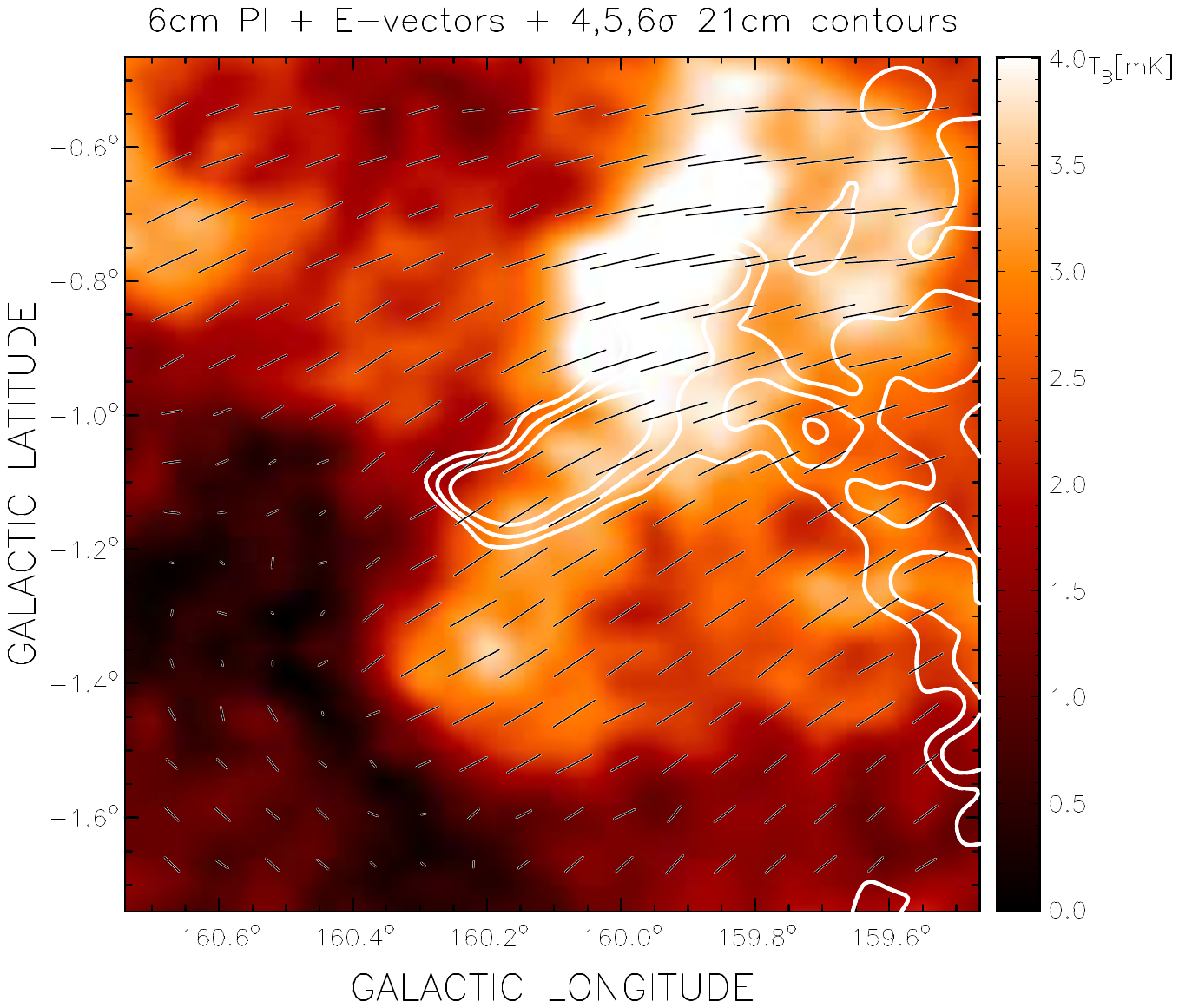}
\includegraphics[bb = 139 241 506 575,height=5.4cm]{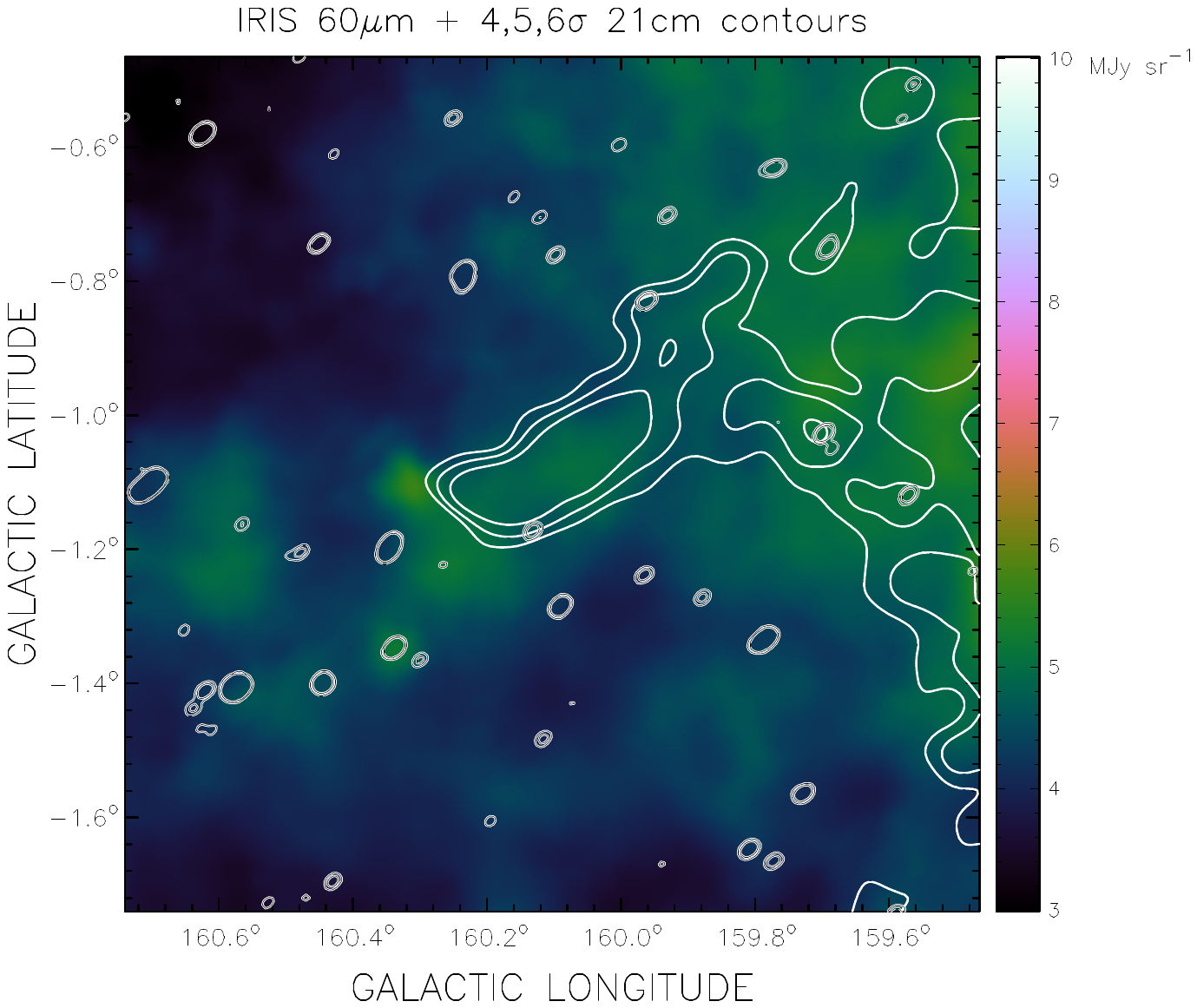}
}
\caption{\label{g160_maps} Same as in Figure~\ref{g128_maps} but for SNR 
candidate G160.1$-$1.1. 
}
\end{figure*}

Total power maps at 4.8~GHz, 2.7~GHz, 1420~MHz, 408~MHz, are 327~MHz are shown 
in Figure~\ref{g160_maps}. In total power, G160.1$-$1.1 has a shell-like shape 
with only one limb detectable, and is best depicted in 1420~MHz total power. 
The long, thin kidney-bean shaped shell is about 36$\arcmin$ long elongated
SE-NW making an angle of 40$\degr$ with the Galactic Plane. G160.1$-$1.1 has 
the steepest spectral index of our five candidates at 
$\alpha_{5}$~=~$-$0.94$\pm$0.10 (see Figure~\ref{g160flux}). The spectral index 
remains at least $-$0.9 whether or not the 92~cm and/or the 6~cm data 
end-points of the spectrum are excluded from the fit. G160.1$-$1.1 is the 
smallest of the five candidates with an angular size of 
36$\arcmin\times$13$\arcmin$.

G160.1$-$1.1 is not significantly polarized at intermediate (1420~MHz) or high 
(4.8~GHz) radio frequencies. The $\lambda$6~cm polarization emission which 
falls partially within the $\lambda$21~cm contours appears to be associated 
with polarized emission beyond the contours, indicating that either the 
polarized emission associated with the object is confused with the larger 
background emission or the polarized emission shown is exclusively associated 
with large-scale, unrelated structure (see map in Figure~\ref{g160_maps}). 
Thus, any polarization evidence for G160.1$-$1.1 is inconclusive.

There is no significant 60~$\mu$m infrared emission (see map in 
Figure~\ref{g160_maps}) within the 1420~MHz contours, and no optical emission 
at all in the field of G160.1$-$1.1. While G160.1$-$1.1 does meet our most 
important criterion of a steep non-thermal radio spectrum, its remarkable 
steepness is hard to explain with the electron energies found in supernova
shocks. Nonetheless, there are other steep-spectrum SNRs \citep[e.g. Cas~A; 
$\alpha=-$0.77][]{gree09} for which G160.1$-$1.1 comes within the
uncertainties. Like Cas~A, G160.1$-$1.1 may be a very young SNR. Based on the 
unmistakeable synchrotron emission spectrum, we thus propose that G160.1$-$1.1 
is a strong SNR candidate. 

\section{Conclusions \& Future}
The purpose of this paper is two-fold. Firstly, we introduce five extended, 
faint discrete objects discovered in the CGPS $\lambda$21~cm dataset and show
evidence (mainly through their radio appearance, spectra and polarization 
properties) that these are to be classified as SNRs. Secondly, we provide a 
basic qualitative interpretation and quantitative observed properties (fluxes 
and sizes) that will aid the community in planning future new observations of 
them at radio wavelengths and especially at other wavelengths (e.g. X-ray, 
optical). The objects were discovered through systematically mining the entire 
CGPS 21~cm dataset after it had been point-source subtracted and smoothed. This 
data processing method increased the signal-to-noise of faint extended emission 
and revealed never-before-seen objects. Their identity as SNRs was then 
ascertained spectrally through comparison with other radio wavelengths (e.g. 
74~cm component of the CGPS; the Sino-German 6~cm polarization survey, etc.), 
and by comparison to other simple criteria (e.g. shape, polarization 
properties, optical and IR appearance).

Found within the outer Galaxy second quadrant (90$\degr\leq\ell\leq$270$\degr$) 
these five newly discovered radio SNRs represent a significant increase 
in the known and suspected Galactic radio SNRs of quadrants II and III of the 
outer Galaxy (see catalogues by \citet{gree09} and \citet{koth06}) which now 
number $\sim$50. In terms of areal density this is 0.65~SNRs~kpc$^{-2}$ which, 
extrapolated across the entire Milky Way stellar disk ($R\sim$14~kpc) suggests 
400 SNRs in the Galaxy, many less than the 1000-2000 expected.
As such, we surmise that many more such faint SNRs can be potentially 
discovered with our approach by mining new high-resolution radio surveys of the 
Milky Way \citep[e.g. GALFACTS][]{tayl10} in order to further our understanding 
of the ISM. 

\begin{acknowledgements}
We wish to thank the referee for constructive help that improved our paper
prior to publication. The Dominion Radio Astrophysical Observatory is a 
national facility operated by the National Research Council Canada. The 
Canadian Galactic Plane Survey is a Canadian project with international 
partners, and is supported by the Natural Sciences and Engineering Research 
Council (NSERC). TF would like to thank Dr. Sean Dougherty (Group Leader, DRAO) 
and the NRCC for their hospitality and support during his sabbatical stay of 
2012-13.
\end{acknowledgements}

\end{document}